\documentclass[preprint2]{aastex63}
\usepackage{graphicx}

\accepted{January 28, 2021}

\shorttitle{MIR HR for HZ planets}
\shortauthors{}

\begin{document}

\title{Detecting Atmospheric Molecules of Temperate Terrestrial Exoplanets using High-Resolution Spectroscopy in the Mid Infrared Domain}

\correspondingauthor{Yuka Fujii}
\email{yuka.fujii.ebihara@gmail.com}

\author[0000-0002-2786-0786]{Yuka Fujii}
\affiliation{National Astronomical Observatory of Japan}
\affiliation{Earth-Life Science Institute, Tokyo Institute of Technology}

\author{Taro Matsuo}
\affiliation{Nagoya University}

\begin{abstract}
Motivated by the development of high-dispersion spectrographs in the mid-infrared (MIR) range, we study their application to the atmospheric characterization of nearby non-transiting temperate terrestrial planets around M-type stars. 
We examine the detectability of CO$_2$, H$_2$O, N$_2$O, and O$_3$ in high-resolution planetary thermal emission spectra at 12-18 $\mu $m assuming an Earth-like profile and a simplified thermal structure. 
The molecular line width of such planets can be comparable to or broader than the Doppler shift due to the planetary orbital motion. Given the likely difficulty in knowing the high-resolution MIR spectrum of the host star with sufficient accuracy, 
we propose to observe the target system at two quadrature phases and extract the differential spectra as the planetary signal. 
In this case, the signals can be substantially suppressed compared with the case where the host star spectrum is perfectly known, as some parts of the spectral features do not remain in the differential spectra. 
Despite this self-subtraction, the CO$_2$ and H$_2$O features of nearby ($\lesssim $ 5~pc) systems with mid-/late-M host stars would be practical with a 6.5-meter-class cryogenic telescope, and orbital inclination could also be constrained for some of them. 
For CO$_2$ and N$_2$O in a 1~bar Earth-like atmosphere, this method would be sensitive when the mixing ratio is 1-10$^3$ ppm. 
The detectability of molecules except O$_3$ is not significantly improved when the spectral resolution is higher than $\mathcal{R}\gtrsim 10,000$, although the constraint on the orbital inclination is improved. 
This study provides some benchmark cases useful for assessing the value of MIR high-resolution spectroscopy in terms of characterization of potentially habitable planets. 

\end{abstract}


\section{Introduction}
\label{s:intro}

Spectroscopy of Earth-sized planets around the so-called habitable zones \citep[HZs; e.g.,][]{Kasting+1993} is one of the key near-future targets in astronomical observations. 
Several approaches have been put forward so far. 
The first major approach is direct (high-contrast) imaging. 
Since the late 20th century, space-based direct imaging possibilities---with coronagraphs in the visible and near-infrared domains, or using nulling interferometers in the mid-infrared (MIR) domain---have been examined and were encapsulated in the proposed missions like TPF-C, TPF-I, and Darwin \citep[e.g.,][]{Levine+2009,Beichman+1999,Leger+1996}. 
These studies are inherited by the ongoing discussions for the future space missions for direct imaging of scattered light of Earth-sized planets (e.g., Starshade with Roman Space Telescope, LUVOIR, HabEx, LIFE). 

In the meanwhile, the successful atmospheric characterization of Hot Jupiters through transmission spectroscopy, eclipse spectroscopy, and detection of phase variations has encouraged its application to smaller Earth-sized planets around late-type stars---this is the second appraoch \citep[see a review by][and references therein]{DemingSeager2017}. 
It has actually been applied to the Hubble Space Telescope (HST) observations of TRAPPIST-1 planets \citep{Gillon+2016,Gillon+2017}, the Earth-sized transiting planets around the habitable zone of an M8-type star, while only the upper limit of atmospheric features have been obtained \citep{deWit+2018,Zhang+2018}.  
The upcoming James Webb Space Telescope (JWST) will be used for more sensitive transit spectroscopy of smaller planets in a wider wavelength range \citep{Beichman+2014,Greene+2016}. 
The high-dispersion spectrographs on the next-generation ground-based telescopes (i.e., extremely large telescopes) may also be used for transit transmission spectroscopy to identify the atmospheric features of Earth-sized exoplanets buried in a number of telluric absorption lines \citep[e.g.,][]{Snellen+2013}. 

The third approach to study potentially habitable exoplanets is to combine high-contrast imaging and high-resolution spectroscopy with these next-generation ground-based telescopes \citep[e.g.,][]{Snellen+2015}. 
Such observations using ground-based telescopes are planned mainly in the near-infrared wavelengths, so as to avoid thermal background that is inevitable for ground-based observations at longer wavelengths, while achieving the spatial resolution necessary for high-contrast imaging. 

\begin{figure}[tb!]
\includegraphics[width=1.0\hsize]{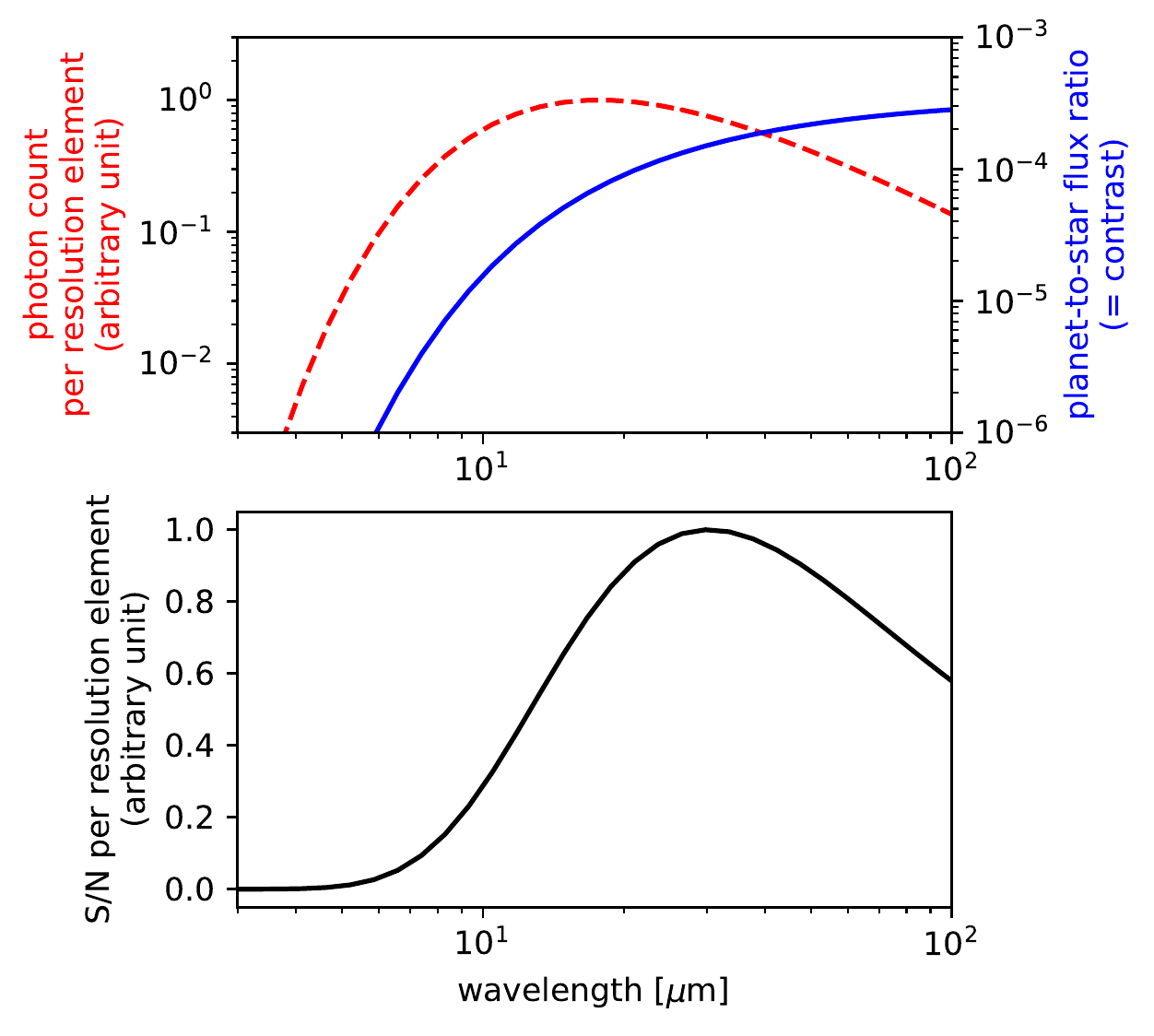}
\caption{Upper panel: A schematic figure of the photon count per wavelength resolution element assuming a black body planet spectrum with 288~K (i.e. $\mathcal{N}_{\rm p}\Delta \lambda$; red dashed) and the planet-to-star flux ratio assuming a black body spectrum mimicking a mid-M host star (i.e. $C$; blue solid). The wavelength dependence of the planet-to-star flux ratio does not depend on the spectral types of the star as long as the stellar spectrum can be approximated by Rayleigh-Jeans law. 
Lower panel: Signal-to-noise ratio per wavelength resolution element as a function of wavelength, estimated by the product of the planet photon count and the planet-to-star flux ratio. }
\label{fig:contrast-photoncount}
\end{figure}

Lastly, the atmospheric features of potentially habitable planets may be identified in the combined spectrum of the star and the planet in the MIR range, without a planetary transit or a starlight-suppressing instrument \citep[e.g.,][]{KreidbergLoeb2016, Snellen+2017}. 
Indeed, the planet-to-star flux ratio of potentially habitable planets is significantly improved in the MIR domain ($\sim $a few tens of parts-per-million or larger) compared with that in the shorter visible domain ($\sim 10^{-7}$ or less) as shown in the upper panel of Figure \ref{fig:contrast-photoncount}, and it is not unrealistic to detect planetary signal in the combined spectra. 
Approximately estimating the wavelength-dependence of the signal-to-noise ratio for the planet signal per wavelength resolution element by $\mathcal{N}_p \Delta \lambda /\sqrt{\mathcal{N_{\star }}\Delta \lambda } = \sqrt{\mathcal{N}_p C \Delta \lambda } = \sqrt{\mathcal{N}_p C \lambda / \mathcal{R}}$ ($\mathcal{N}_p$ and $\mathcal{N}_{\star }$ are the photon spectrum of the planet and the star, respectively, $\Delta \lambda$ is the wavelength width of the resolution element, $C$ is the planet-to-star flux ratio, i.e., $C\equiv \mathcal{N}_p / \mathcal{N}_{\star }$, and $\mathcal{R}$ is the fixed spectral resolution), the MIR domain, specifically around 13-100~$\mu $m, is most useful (the lower panel of Figure \ref{fig:contrast-photoncount}). 
The unique advantages of this MIR approach are that (1) unlike high-contrast technique it does not require specialized instruments to occult stellar flux, and that (2) unlike transmission or eclipse spectroscopy it can be applied to both transiting and non-transiting planets. 
The latter is critical to increase the number of targets, as the transit probability of habitable-zone planets is up to about 5\%. 

Along this line, \citet{KreidbergLoeb2016} proposed low-resolution spectroscopy of Proxima Centauri systems to try to detect 9.6 $\mu $m O$_3$ features originated from the atmosphere of Proxima Centauri b, the nearest non-transiting potentially habitable planets  \citep{Anglada-Escude+2016}. 
 
\citet{Snellen+2017} proposed that medium-resolution spectroscopy (MRS) in the MIR range can be used to identify high-frequency features due to planetary atmospheric molecules in the combined spectra, and they estimated that CO$_2$ features of Proxima Centauri b would be detected after 5 days of observations with MRS mode of JWST/MIRI. 
Although these are encouraging possibilities, such observations rely on the precise knowledge of the stellar spectrum as well as the sensitivity of the detector elements. 

While the above studies have in mind the upcoming JWST that have only low-resolution and medium-resolution capabilities in the MIR, technologies for high-dispersion spectrograph in the MIR have been recently developed for future cryogenic space telescopes, including SPICA \citep[e.g.,][]{Sarugaku+2012}. 
The technologies are expected to largely reduce the size of the MIR spectrograph and enhance its throughput. 
In the further future, Origins Space Telescope (OST) is also contemplating high-resolution spectrograph with a larger aperture \citep{Sakon+2018}. 
With high-resolution spectroscopy ($\mathcal{R} \sim 30,000$), not only the fine structures of absorption bands of the planetary spectrum can be identified, but also the Doppler shift of a HZ planet can be resolved (unless its orbit is close to face-on). 
Thanks to the Doppler-shift, it is possible to identify planetary features even without the assumption for the stellar spectrum, and in theory the orbital inclination can also be constrained. 
Although these new opportunities will be useful for characterizing atmospheres of potentially habitable planets, its potential remains largely unexplored yet.

In this paper, we examine the utility of High-Resolution Spectroscopy (HRS) in the MIR for characterizing atmospheres of nearby potentially habitable planets. 
In Section \ref{s:theoretical_spectra}, we begin by demonstrating the MIR spectral features of potentially habitable planets using a simplified 1-dimensional atmospheric model. 
We then move to detectability assessment in Section \ref{s:detectability}. We propose to observe the target system at two quadrature phases, and discuss how this method could constrain the presence of molecular features and the orbital inclination under the realistic assumptions on the observational noise, highlighting the difference between the two analysis strategies that reflect the possible uncertainty in the prior knowledge of the stellar spectrum. 
The dependence on the spectral resolution, bandpass, and the abundance of molecules are examined in Section \ref{s:Detecability_Resolution_Abundance}. 
Section \ref{s:discussion} discusses the range of the targets and the synergies with other techniques to characterize potentially habitable planets (Section \ref{ss:targets}), the effects of other noise (Section \ref{ss:other_noise}), the effect of different thermal profiles of the planet (\S\ref{ss:dependence_on_atmosphere}), and the comparison with the previously studied medium-resolution spectroscopy (\S\ref{ss:comp_MRS}). 
Finally, Section \ref{s:summary} summarizes our results.

\section{Atmospheric features in MIR high-resolution spectra}
\label{s:theoretical_spectra}

In this section, we model high-resolution planetary spectra of an Earth-like planet and discuss the characteristics of the spectral features of molecules of interest.
These modeled spectra will be used as input for the detectability analyses in Section \ref{s:detectability}. 

\subsection{Model Atmosphere}
\label{ss:method_spectra}

\begin{figure}[tb!]
\includegraphics[width=1.0\hsize]{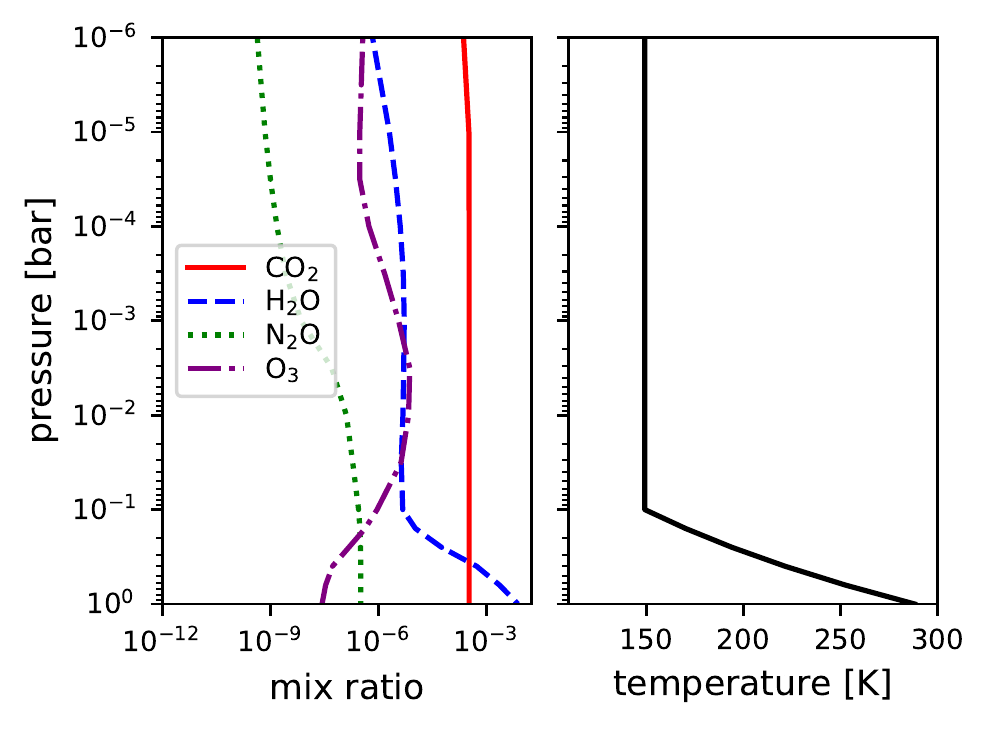}
\caption{The assumed vertical profiles of the mixing ratios of the molecules considered (left) and temperature (right). The mixing ratio of molecules are based on ``US standard'' model.  The temperature profile is determined by the dry adiabatic lapse rate (9.8~K/km) in the lower atmospheres below 0.1~bar, above which an isothermal profile is assumed. }
\label{fig:simple_TP_profile}
\end{figure}

Our model atmosphere is based on Earth's atmosphere except for the temperature profile. 
We consider four molecules that are present in Earth's atmosphere and are radiatively active in the mid infrared range: CO$_2$, H$_2$O, N$_2$O, and O$_3$.
The vertical profiles of the mixing ratios of these molecules are taken from the ``US standard'' model and are shown in the left panel of Figure \ref{fig:simple_TP_profile}.  
The surface temperature is set at 288~K, again referring to the ``US standard'' model. 
The vertical temperature profile is, however, replaced by a simplified one comprised of a troposphere with a constant lapse rate ($\Gamma =g/C_p$ where $g$ is the gravity and $C_p$ is the specific heat capacity) and an isothermal stratosphere. 
The removal of stratospheric thermal inversion is motivated by the fact that O$_3$, even if it exists, does not lead to a strong thermal inversion under the irradiation of M-type stars due to the reduced near-UV flux. 
The effects of the atmospheric profile on the detectability of molecules are discussed in Section \ref{sss:thermal_inversion}. 
For simplicity, the effects of clouds are ignored.

The surface pressure is fixed at 1~bar, and the tropopause is set at 0.1~bar. 
This tropopause pressure is consistent with the observations of Solar system planets \citep[e.g.,][]{RobinsonCatling2014} and 3D climate simulations for habitable planets around M-type stars \citep[e.g.,][]{Fujii+2017}.

\begin{table}[]
\centering
\begin{tabular}{lcc}\hline \hline
description          & symbol          & value\\ \hline
surface pressure     & $P_{\rm surf} $ & 1~bar \\
surface temperature  & $T_{\rm surf} $ & 288~K \\
tropopause & $P_{\rm tp} $ & 0.1~bar \\ 
surface gravity & $g $ & 9.8~m/s$^2$ \\
tropospheric lapse rate & $\Gamma (=g/C_p)$ & 9.8~K/km \\ \hline
\end{tabular}
\caption{Assumptions for atmospheric profiles. }
\label{tbl:parameters}
\end{table}

Given a vertical profile, the top-of-atmosphere outgoing radiance at wavelength $\lambda $ and the cosine of the zenith angle $\mu $, $L(\lambda , \mu )$, is computed by
\begin{eqnarray}
L(\lambda, \mu ) &=& \int B( T; \lambda ) \exp(-\tau /\mu ) d\tau/\mu  \label{eq:thermalemission}\\
\tau &\equiv& \int _z ^{\infty } k[T(z),P(z)] x(z) n(z) dz \\
&=& \int _0^p k[T(P),P] \, x(P) \frac{dP}{\mu_{\rm atm} g } \label{eq:tau}
\end{eqnarray}
assuming a no-scattering atmosphere. 

The cross sections of molecules are based on HITRAN2016 \citep{Gordon+2017} and the lines are broadened by the Voigt functions using the algorithm of \citet{Zaghloul+2011} with the partition functions  adopted from HAPI program \citep{Kochanov+2016}. 
We impose the cut-off of the line wings at 100 cm$^{-1}$ apart from the line centers. 
We do not include the continuum absorption (e.g., H$_2$O continuum), as it does not significantly affect the detectability of high-resolution features. 

For the efficient evaluation of equation (\ref{eq:thermalemission}) at a large number of the wavelength grid points, we employed GPU computation through PyCUDA. 
The integral in equation (\ref{eq:tau}) is performed with 50 equi-distributed points in $\log P$ space  and that in equation (\ref{eq:thermalemission}) with 50 equi-distributed points in $\log \tau $ space (from $10^{-2}$ to $10^2$), using the trapezoidal rule. 

Thermal emission spectrum is calculated with the opacities of all four molecules included. 
In addition, we also calculate the spectra with the opacity of only one molecule turned on, in order to discuss the spectral characteristics and detectability of individual molecules in isolation. 

To obtain the total thermal emission from the planet, equation (\ref{eq:thermalemission}) needs to be integrated over the planetary disk. 
With several trials, we find that the disk-averaged radiance that integrates equation (\ref{eq:thermalemission}) over different $\mu $ with the weight of the projected area is close to the radiance at $\mu =0.6$. 
Thus, we represent the disk-averaged radiance by the radiance with $\mu =0.6$. 
Doppler broadening due to planet rotation is not included; HZ planets around M-type stars are likely to be tidally locked, which means the rotational velocity of $\sim 10$--100 m/s, and the corresponding to the wavelength shift, $\sim 10^{-6}\;\mu $m, is sufficiently smaller than the wavelength resolution elements considered in this paper.

\subsection{Model Spectra and Their Characteristics}
\label{ss:result_spectra}

\begin{figure*}[tb!]
\includegraphics[width=1.0\hsize]{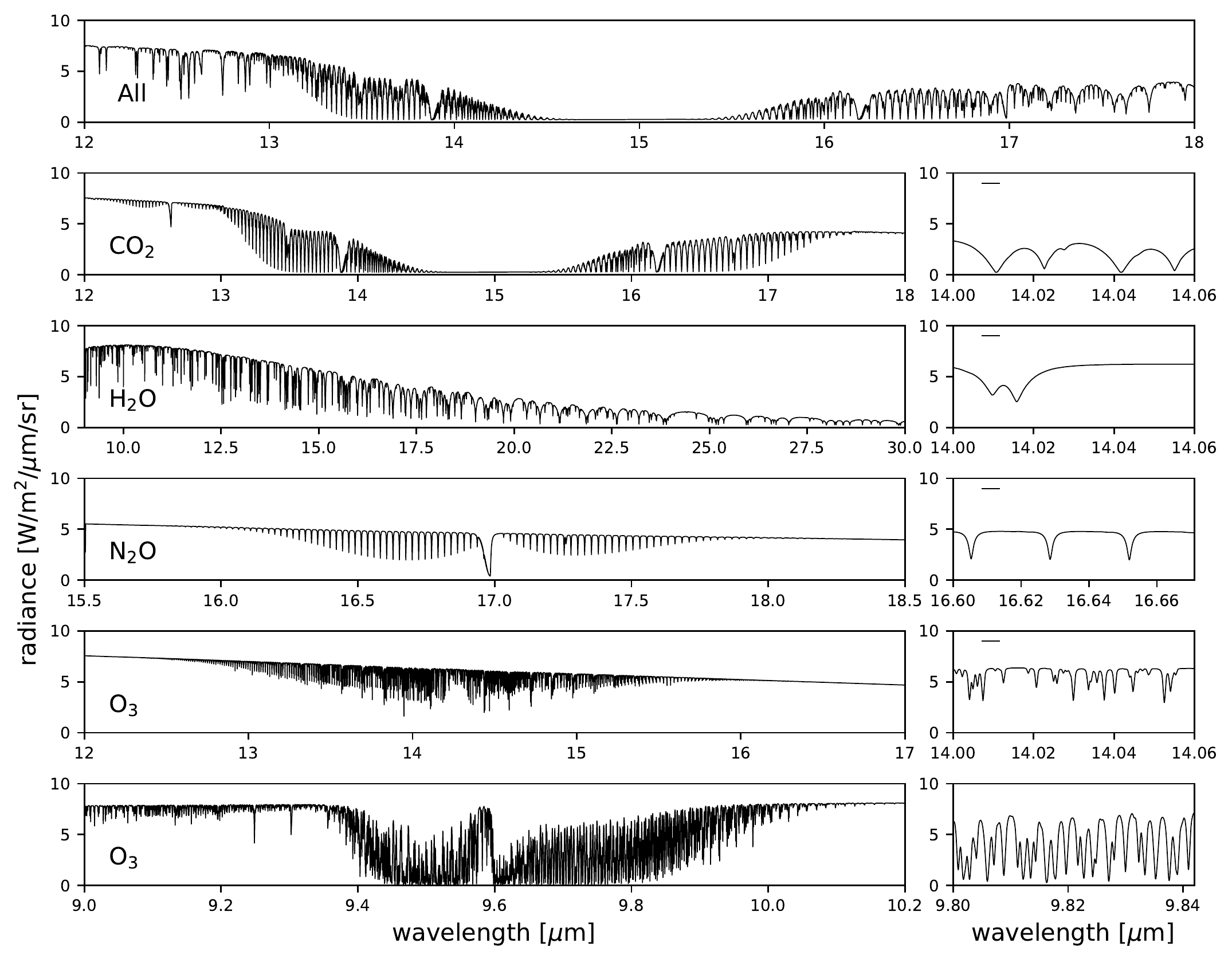}
\caption{High-resolution ($\mathcal{R}=30,000$) features of molecules in thermal emission spectra assuming the vertical temperature profile and molecular mixing ratio shown in Figure \ref{fig:simple_TP_profile}. In the uppermost panel the opacities of all molecules are included, while in the lower panels the opacity of one molecule is assumed at a time. The left panels show the broadband spectra while the right panels show the zoom-in where the horizontal bars approximately indicate the maximum range of the Doppler shift (corresponding to $\sim $100~km/s).}
\label{fig:HRspectra}
\end{figure*}

Figure \ref{fig:HRspectra} presents the major features in high-resolution spectra ($\mathcal{R}=30,000$) assuming the temperature profile shown in Figure \ref{fig:simple_TP_profile}. 
The uppermost panel includes the opacities of all four molecules, while the lower panels turn off the opacities of molecules other than the indicated one, showing more clearly the spectral features of individual molecules. 

CO$_2$ has the prominent rotational-vibrational features around 15~$\mu $m associated with the fundamental bending mode ($\nu _2$).  
The strongest lines on the P- and R-branches are separated by $\sim 0.04~ \mu $m, corresponding to the rotational energy $\sim 1.5$ cm$^{-1}$, well resolved with $\mathcal{R}=30,000$. 
On the other hand, H$_2$O has rotational lines that broadly spread in the MIR range. 
The 17~$\mu $m N$_2$O band is mainly due to the  bending mode ($\nu _2$) and has a structure similar to the 15~$\mu $m CO$_2$ band. 
Among these, CO$_2$ largely contribute to shaping the overall shape of the all-included spectrum, although the minor features due to H$_2$O and N$_2$O can also be seen. 

Compared to these bands, the famous 9.7 $\mu $m O$_3$ band with the overlapping fundamental vibration modes (symmetric $\nu _1$ and asymmetric $\nu _3$) is densely populated with lines. 
A weaker O$_3$ band exists around 14.5 $\mu $m, corresponding to the bending mode, which also features clustered lines in a narrow bandpass. 
The latter is largely masked by CO$_2$ in Earth's thermal emission spectra. We find that this band start to kick in when CO$_2$ is smaller than $\sim 1$~ppm. 
Do such CO$_2$-poor atmosphere exist? It is possible, as CO$_2$ abundance in general depends on the carbon cycles, as well as how much the planet acquires and retains carbon. 
Planets which develops higher weathering rate may turn into CO$_2$-poor worlds \citep[e.g.,][]{Nakayama+2019}. 
In the absence of CO$_2$ features, 14.5~$\mu $m O$_3$ band may be more useful than the 9.7 $\mu $m band, due to the better intrinsic detectability (Fig. \ref{fig:contrast-photoncount}); see more discussions in Section \ref{ss:choise_of_bandpass}.

Here we highlight a characteristics of these high-resolution features---the relative width of the lines. 
As indicated in the right panel of Figure \ref{fig:HRspectra}, the broad lines extend beyond the wavelength resolution (5$\times 10^{-4}$~$\mu $m at 15 $\mu $m) and the width of the Doppler shift ($\sim $0.005 $\mu $m at 15 $\mu $m; horizontal bars). 
This is partly due to the intrinsic broadening of the lines. 
The half width of the Lorentz profile (collisional broadening), which is important at pressures higher than $\sim 0.01$~bar, is approximately constant in {\it wavenumber} for a given set of pressure and temperature, i.e., the width {\it relative to the wavelength} is wider at longer wavelengths. 
At 1~bar and 288~K, it is approximately 0.1~cm$^{-1}$, corresponding to 0.002~$\mu $m. 
The line shape also depends on the abundance of the molecule and the vertical temperature gradient. 
Lines become broad if the opacity is so large that the atmosphere becomes optically thick at the far wings; this happens for CO$_2$ and H$_2$O with our model atmosphere.
The broad lines have notable influence on the analysis, as we will see in the next section. 

\section{Detectability}
\label{s:detectability}

In this section, we examine the detectability of MIR molecular features of temperate Earth-sized planets presented in Section \ref{s:theoretical_spectra}, assuming a high-dispersion spectrograph mounted on a cryogenic telescope. 
We try to identify molecular features in the combined spectrum of the host star and the planet. 
The assumptions for mock observations and noise estimate are given in Section \ref{ss:mock_observations}, which is followed by the description of our analysis procedures in Section \ref{ss:analysis}. 
The resultant constraints on the contrast and the orbital inclination are discussed in Section \ref{ss:results}.

\subsection{Mock observation}
\label{ss:mock_observations}

\subsubsection{Targets}
\label{sss:targets}

\begin{table}[bt!]
\centering
\caption{Assumptions for the host star and the planetary orbit.}
\begin{tabular}{lcc}\hline \hline
description    & M5 star & M8 star \\ \hline
star radius ($R_{\star } $) & 0.14 $R_{\odot }$ & 0.10$R_{\odot }$ \\
star temperature ($T_{\star} $) & 3000~K & 2500~K \\
planet/star flux ratio\tablenotemark{a} & $\sim 70$~ppm & $\sim 200$~ppm \\
planet orbital radius  & 0.0485~au & 0.0146~au  \\ 
planet orbital period  & 11.26 days & 2.27 days \\
planet orbital velocity  & 46.83 km/s & 69.70 km/s \\ \hline
\end{tabular}
\tablenotetext{a}{Evaluated at 15$\mu $m, assuming a black body spectrum with 288~K for the planet spectrum. }
\label{tbl:systemparameters}
\end{table}

The prime target of this study is temperate rocky planets around M-type stars, because those around earlier-type stars have too large planet-to-star flux ratio for planetary signals to be detectable in a reasonable amount of time. 
We initially considered three types of host star: early-, mid- and late-M stars. 
However, the typical planet-to-star flux ratio with early-M stars is smaller than 10~ppm, making it extremely challenging to detect planetary features.  
Therefore, we focus on mid-M and late-M stars. 

Our fiducial model assumes an M5 or M8 star at 5 parsecs.
The assumed host star and planetary parameters are summarized in Table \ref{tbl:systemparameters}. 
Additionally, the case study for Proxima Centauri b is also presented, where the host star properties are the same as those of M5 star in Table \ref{tbl:systemparameters}, the distance is set to 1.3~parsecs, and the planetary radius is assumed to be 1.1 Earth radius \citep[e.g.,][]{Snellen+2017}. 

The inclination of the planetary orbit, which controls the amplitude of the planetary line-of-sight velocity, is fixed at $60^{\circ }$. 

The host star spectra are taken from the BT-Settl model \citep{Allard+2012} with the corresponding effective temperatures, while assuming $\log g=5.0$ and [Fe/H]$=0.0$. 
This is an update from the previous studies on the detectability of spectral features of Proxima Centauri b where the black-body spectrum is assumed \citep{Snellen+2017, KreidbergLoeb2016}.

\subsubsection{Configuration}
\label{sss:config}

\begin{figure}[htb!]
\includegraphics[width=1.0\hsize]{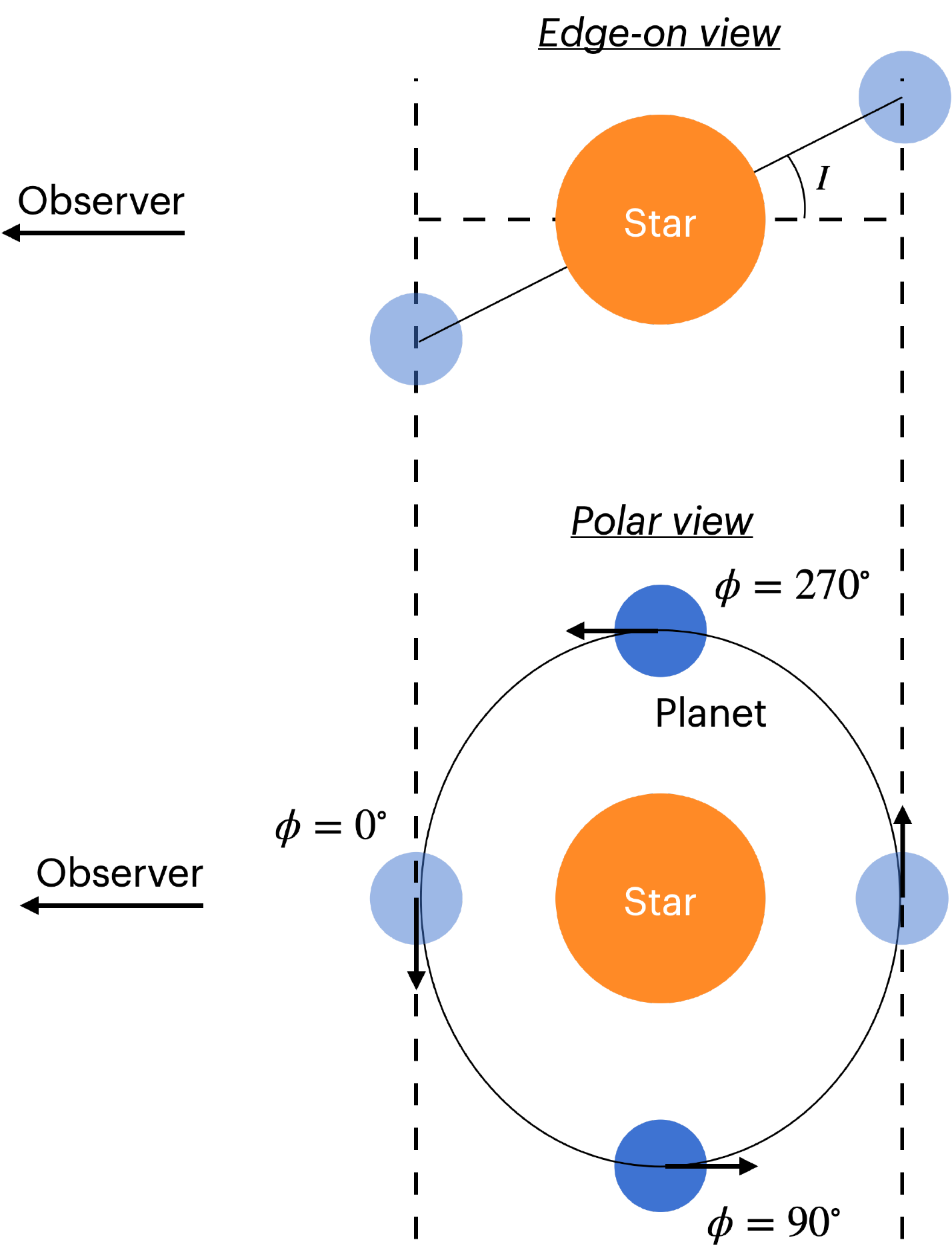}
\caption{Observational configuration showing the geometrical parameters. The data near $\phi =90^{\circ }$ and $\phi =270^{\circ }$ are used in this study. }
\label{fig:schematics}
\end{figure}

Figure \ref{fig:schematics} illustrates the observational configuration. 
We assume that the mock observations are carried out when the planet is near $\phi =90^{\circ }$ and near $\phi =270^{\circ }$, where $\phi $ is the orbital longitude (Figure \ref{fig:schematics}). 
This is because the orbital inclination would be constrained best when the data cover the orbital phases where the radial velocity changes the most. 

For a planet around a M5 (M8) star, mock observations are continued for 1 (0.25) day centered at $\phi = 90^{\circ }$ and for another 1 (0.25) day centered at $\phi = 270^{\circ }$, which cover approximately $32^{\circ }$ ($40^{\circ }$) on each side. 
These two cadences are repeated until the parameters are constrained, and the total integration time is recorded.

\begin{figure}[htb!]
\includegraphics[width=1.0\hsize]{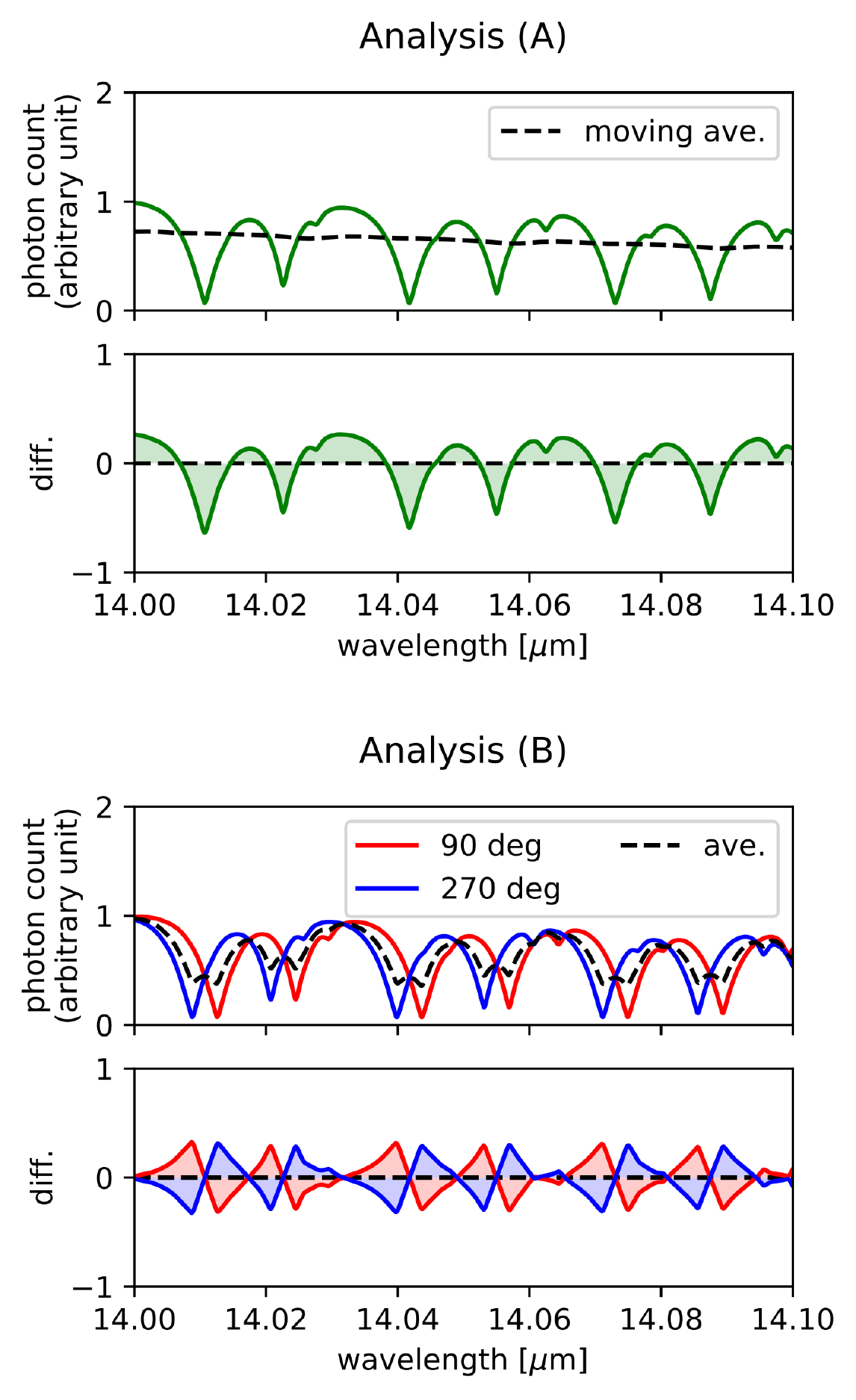}
\caption{Schematic figures showing the signals to be detected. Analysis (A) (left) utilizes the high-frequency features of the spectra extracted by subtracting the moving average from the spectrum, while Analysis (B) (right) utilizes the difference of the Doppler shifted spectra from the average spectra.}
\label{fig:schematics2}
\end{figure}

\subsubsection{Instruments and observational configurations}
\label{sss:obs_strategy}

Mock observations are carried out with a high-resolution spectrograph at 12-18 $\mu $m with the resolving power of $\mathcal{R}=30,000$. 
This bandpass and resolving power are motivated by the latest specification of SPICA's SMI/HR ($\mathcal{R}\sim 33,000$ in 12-18 $\mu $m) \citep[e.g.,][]{Kaneda+2018} and by the expected resolution of OST/MISC  \citep[e.g.,][]{Sakon+2018}. 
The dependences on the spectral resolution and the bandpass are examined in Section \ref{s:Detecability_Resolution_Abundance}.

The total throughput including the quantum efficiency of the detector is assumed to be 0.2. 
We note that the throughput of the optical system of existing high-resolution spectrographs installed on ground-based telescopes have reached approximately 60\% \citep[e.g.,][]{Ikeda+2016,Ikeda+2018}, and the throughput of future instruments could be higher than what we assume here. 
It is trivial to scale our results by throughput; see Section \ref{sss:scaling}. 

We collect data every 1800 sec (i.e., exposure time is assumed to be 1800 sec). 
Due to the change of the planetary radial velocity, the planetary spectrum is Doppler shifted relative to the host star spectrum on the detector plane. 
Over the course of the planetary orbital motion, the planetary spectrum moves beyond the resolution elements unless the orbital inclination is close to zero (i.e., face-on orbit), as the radial velocity amplitudes  shown in Table \ref{tbl:systemparameters} (and the assumed orbital inclination of $60^{\circ }$) are larger than the velocity corresponding to the resolution element, $c/\mathcal{R}\sim 10$~km/s. )

\subsubsection{Signal and noise}
\label{sss:noise}

The total photoelectron count that the detector receives is the summation of the planetary light ($\mathcal{N}_{{\rm p}}$), stellar light ($\mathcal{N}_{\star}$), zodiacal light ($\mathcal{N}_{{\rm zodi}}$), thermal background of the telescope ($\mathcal{N}_{{\rm tele}}$), and the dark current ($\mathcal{N}_{{\rm dark}}$). 
We assume that the contribution from the zodical light and the dark current are perfectly subtracted through a post-processing. 
This leaves $\mathcal{N_{\rm total}} = \mathcal{N}_{{\star }} + \mathcal{N}_{{\rm p}}$ alone as a signal.

\begin{table}[tb!]
\centering
\begin{tabular}{lcc}\hline \hline
description          & symbol          & value\\ \hline
spectral resolution & $\mathcal{R}$ & 30,000 \\
telescope diameter     & $D$ & 6.5~m \\
total throughput    & $\xi $ & 0.15 \\
distance to target  & $d$ & 5~pc \\
planetary radius & $R_{\rm p}$ & $R_{\oplus }(=6.371\times 10^6$~m) \\
exposure time  & $\tau _{\rm exp}$ & 1800 sec \\
\hline
\end{tabular}
\caption{Fiducial values for observational parameters. The scaling of the required integration time by these parameters are given in equation (\ref{eq:scaling}).}
\label{tbl:parameters}
\end{table}

The shot noise from all of these factors contributes to the observational noise, although the shot noise due to the planetary flux is negligible compared to that from the stellar flux. 
An additional factor that is taken into account is the read noise. 
The systematic noises such as a fringe of the detector are assumed to be perfectly removed; the effect of systematic noise is discussed in Section \ref{sss:systematic_noise}. 
These assumptions imply the Gaussian random noise with the following standard deviation for $j$-th wavelength element at wavelength $\lambda _j$:
\begin{eqnarray}
\sigma^2_{{\rm photon},j} =&& \mathcal{N}_{{\rm star},\,j} ( 1 + \xi b_j(T_{\star }) ) + \mathcal{N}_{{\rm zodi},\,j} \nonumber \\
&&+ \mathcal{N}_{{\rm tele},\,j}+ \mathcal{N}_{{\rm dark},\,j} + \sigma_{\rm read}^2
\label{eq:sigma}
\end{eqnarray}
where $\sigma_{\rm read}^2 $ represents the read noise, $b(T)$ is the Bose factor, and $\xi $ is the total throughput. 

The Bose factor, $b(T)$, is to take account of the sub-Poissonian nature of the photon count statistics at $h\nu \ll kT$ \citep[e.g.,][]{Boyd1982}:
\begin{equation}
    b_j(T) = \frac{1}{\exp \left( \frac{hc}{\lambda _j kT} \right) - 1 }
\end{equation}
For MIR observations ($\lambda \sim 15\,\mu $m) of  M-type stars ($\sim 3000$~K), $b \sim 2.7$. 
Multiplied by the throughput ($\xi =0.2$), the Bose factor increases the variance of the photon count by about 50\%. 
The Bose factor for the zodiacal light is not included due to the low temperature. 

\begin{figure}[bt!]
\includegraphics[width=1.0\hsize]{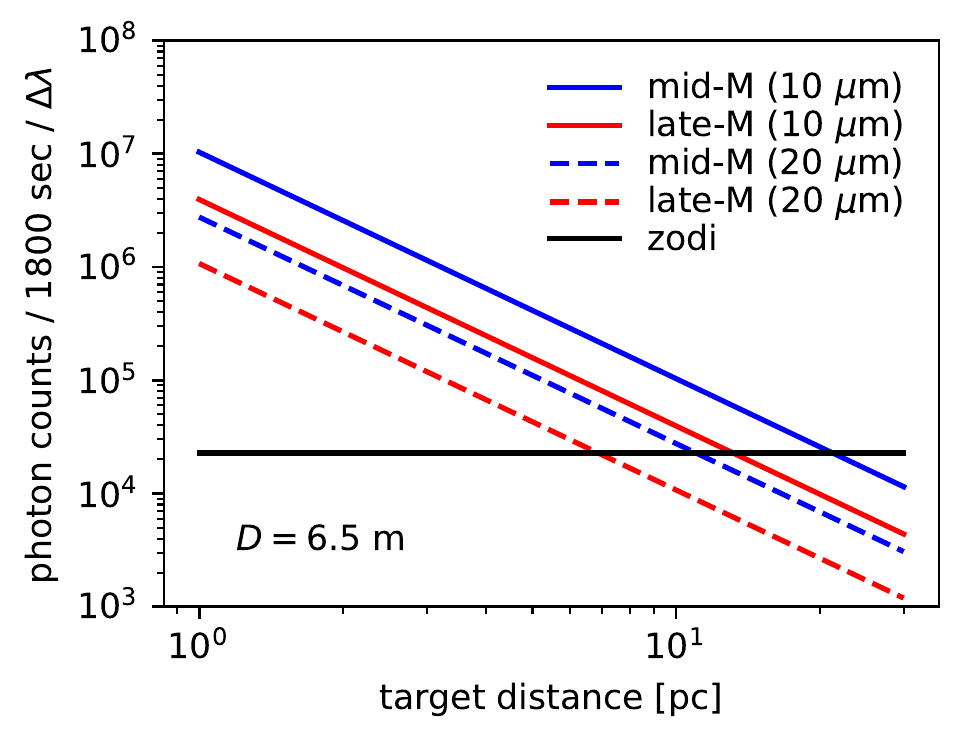}
\caption{Photon count of the zodiacal light (black) and the starlight (colors) at wavelength 10~$\mu $m (solid line) and at 20~$\mu $m (dashed line), per exposure time (1800~sec) per wavelength resolution element ($\lambda / \mathcal{R}$). These numbers can also be compared with dark current ($360e^{-1}$/exposure) read noise, $14\times 4 e^{-}$/exposure.}
\label{fig:photon_star_zodi}
\end{figure}

In the fiducial cases studied in this paper, the dominant noise source is the stellar flux. 
The contributions of these factors in general cases are presented in Figure \ref{fig:photon_star_zodi}, as a function of the spectral type of the star and the distance to the target.

Each term in equation (\ref{eq:sigma}) is expanded below.

\paragraph{Planet/star spectra}

The photoelectron counts of the planet and the host star per exposure are simply:
\begin{eqnarray}
    \mathcal{N}_{{\rm p}} (\lambda ) &=& \mathcal{F} L_{\rm p}(\lambda ) \cdot \pi R_{\rm p}^2 \left( \frac{1}{d} \right)^2 \\
    \mathcal{N}_{\star} (\lambda ) &=& \mathcal{F} L_{\star} (\lambda ) \cdot \pi R_{\star }^2 \left( \frac{1}{d} \right)^2 \\
    \mathcal{F} &\equiv & \pi \left( \frac{D}{2} \right)^2 \xi \tau_{\rm exp} \frac{\lambda }{\mathcal{R}} 
\end{eqnarray}
where $L_{\rm p}$ and $L_{\star}$ represent the radiance of the planet and star, respectively, while $R_{\rm p}$ and $R_{\star }$ represent the radius of the planet and the star, respectively. 
The meanings of other parameters are summarized in Table \ref{tbl:parameters} together with our fiducial values. 
Here $\mathcal{F}$ is used to denote the common factor related to the observational configuration.

\paragraph{Zodiacal light}
\label{sss:starlight}

The zodiacal flux of the Solar System is set to 15~MJy/sr \citep{Glasse+2015} and is assumed to be constant over the observing wavelength range. 
The zodiacal light in the target system is not included in our simulation because it is much fainter than the other sources. 
Thus, 
\begin{equation}
\mathcal{N}_{{\rm zodi}} = \mathcal{F} f_{\rm zodi} \cdot \pi ( \theta_{\rm aperture})^{2}
\end{equation}
where $f_{\rm zodi}$ is the zodiacal flux per wavelength per steradian (rather than per frequency as represented in Jansky per steradian) and $\theta_{\rm aperture}$ denotes the aperture radius. 
The aperture radius should be determined by the balance between the shot noise of the thermal background and the systematic noise \citep{Matsuo+2018}. 
Based on the analytical formulation on the relation between the aperture radius and systematic noise \citep{Itoh+2017}, the aperture radius is set to 1.85 arcsecond, corresponding to 4 times the diffraction limit at 15 $\micron$ such that the systematic noise is reduced down to 100 ppm under a condition that the pointing jitter of the Origins Space Telescope is 22 milli-arcsecond (RMS), corresponding to approximately 0.05 $\lambda / D$ at 15 \micron{}  \citep{Leisawitz+2018}.

\paragraph{Telescope background}
\label{sss:tele}
The thermal light and stray light from a telescope assembly also contributes to the background light. For the JWST MIRI, the telescope background can be approximately fitted by a combination of several blackbody radiations with temperatures ranging from 50 to 70~K \citep{Glasse+2015}. When the telescope is cooled down to below 10~K, as planned for SPICA and OST, and the telescope background becomes negligible at $<$ 100 $\mu$m.

\paragraph{Dark current}
\label{sss:darkcurrent}

The dark current is assumed to be 0.2 $e^-$~s$^{-1}$ based on the performance of the Si:As detector for the JWST MIRI \citep{Rieke+2015}, but this never becomes of relative importance in our study.

\paragraph{Read noise}
\label{sss:readnoise}

The read noise is assumed to be 14 $e^-$/read assuming the Fowler-eight sampling \citep{Rieke+2015}. 
The loss time due to the readout is not considered in the paper. 
Each resolution element is sampled by 4 pixels. 
With these assumptions, read noise never becomes dominant for the exposure time of 1800 sec. 
We note that 1800 sec is longer than the typical values for JWST. 
The reduction in the exposure time can lead to a significant contribution from read noise. 

\subsection{Analysis}
\label{ss:analysis}

Because the planetary signal is a tiny portion in the total spectrum, it is critical to subtract the stellar spectra precisely from the total spectrum. 
Previous studies discussing the detectability of atmospheric features of non-transiting habitable planets assumed that the host star spectrum is determined precisely \citep[e.g.,][]{KreidbergLoeb2016,Snellen+2017}. 
The analysis based on this optimistic assumption is denoted by Analysis (A) and is described in Section \ref{sss:analysis_optimistic}. 

\begin{figure}[bt!]
\includegraphics[width=1.0\hsize]{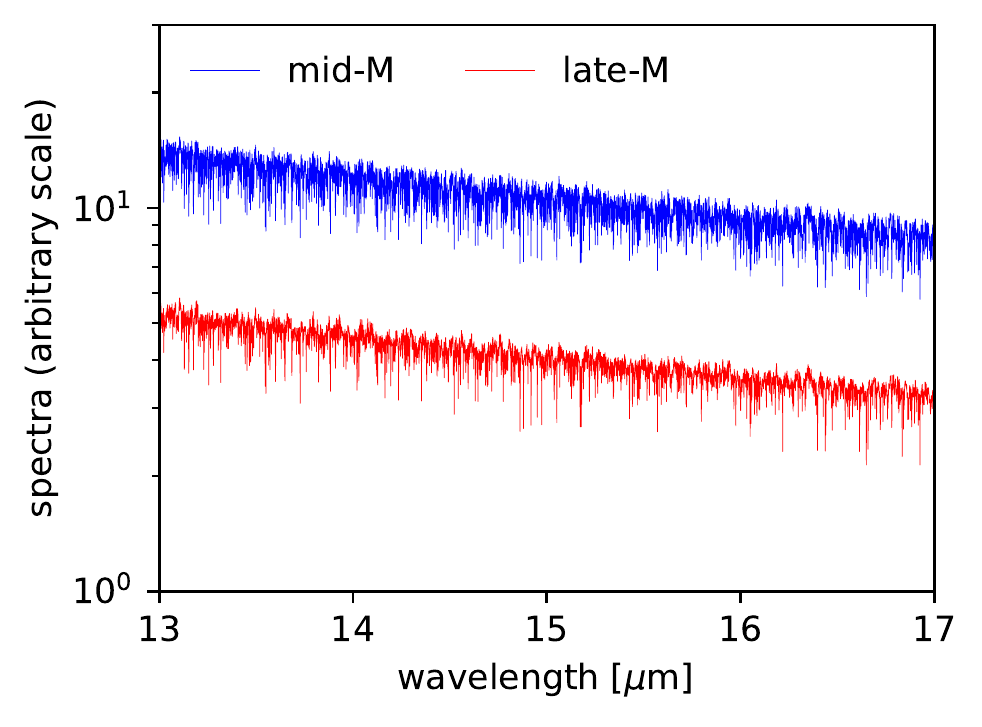}
\caption{MIR spectra based on BT-Settl model \citep{Allard+2012}.}
\label{fig:star_MIR}
\end{figure}

However, M-type stars are rich in spectral features as shown in Figure \ref{fig:star_MIR} and its MIR stellar spectrum may not be determined to a satisfactory accuracy, due to the inaccurate line list and to the uncertainties in the stellar atmospheric structure. 
Indeed, the models of star spectra do not fit the near-infrared observed spectrum of M-type stars to the precision of observed data \citep[e.g.,][]{Zhang+2018,Wakeford+2019}.
In the case where we do not have any prior knowledge about the stellar spectrum, we have to focus on the Doppler-shifted components in the combined spectrum of the star and the planet to extract planetary one. 
Such an analysis is denoted by Analysis (B) and is described in Section \ref{sss:analysis_pessimistic}. 

The planetary signals to be fitted in the case of Analysis (A) and (B) are  illustrated in Figure \ref{fig:schematics2}.
Importantly, as described in more detail below, the signal in Analysis (B) is smaller than that in Analysis (A) because some fraction of the spectral features are cancelled out. 

\subsubsection{Analysis (A): with a well-modeled stellar spectrum}
\label{sss:analysis_optimistic}

In the first analysis, we follow the procedure of \citet{Snellen+2017}, assuming that the fine structure of the stellar spectrum can be modeled precisely. 
Even with this assumption, we cannot separate the host star and the planet spectra from observations alone, so we take the following steps:
\begin{enumerate}
    \item The model stellar spectrum is fitted to the data (that is the combined spectrum of the star and the planet), by varying the scaling of the spectrum as well as the stellar parameters. 
    \item The best-fit stellar spectrum is subtracted from the data. 
    \item The residual spectrum is corrected by subtracting its moving-average. 
    \item The model planet spectrum, which is also corrected through the moving-average subtraction (the bottom panel of Figure \ref{fig:schematics2}), is fitted to the corrected residual spectra. 
\end{enumerate}

In Step 1, we use the input BT-Settl model and fit for the absolute scale.  
The best-fit scale is slightly larger than the input stellar model, due to the contribution from the planet spectrum. 
In other words, some fraction of the planetary spectrum is subtracted in Step 2. 
This results in a trend in the residual spectrum, which is corrected in Step 3. 

Finally in Step 4, the high-frequency features are fitted by the model planet spectrum. 
For simplicity, our fitting model is the same as the input model, similar to \citet{Snellen+2017}, and we consider two fitting parameters: the planet-to-star contrast $C$ and the orbital inclination $I$. 
The parameter estimate is based on the posterior probability, $\mathcal{P} (c,i|\{\mathcal{N}_i (t_k) \})$:
\begin{equation}
    \mathcal{P} (C,I|\{\mathcal{N}_{{\rm res}, jk} \}) = \mathcal{L} ( \{\mathcal{N}_{{\rm res}, jk} \}|C, I ) \Pi(C) \Pi(I) \label{eq:postprob}
\end{equation}
The $\mathcal{N}_{{\rm res},jk}$ represents the corrected residual spectra at $j$-th wavelength elements and $k$-th observation epoch. 
The likelihood is simply:
\begin{eqnarray}
&& \mathcal{L} ( \{\mathcal{N}_{{\rm res}, jk}|c, i ) \propto \exp \left( - \frac{\chi^2}{2} \right) \label{eq:likelihood} \\
&& \chi^2 \equiv   \sum _{\{j, k\}} \left( \frac{\mathcal{N}_{{\rm res},jk}^{\rm obs} - C  \mathcal{N}_{{\rm res},jk}^{\rm theory} (I) }{\sigma _{j,k}} \right)^2. \label{eq:chi2}
\end{eqnarray}
We assume a flat prior for contrast $C$ between 0 and 10, and a flat prior for the orbital inclination $I$ between 0 to 90 degree; $\Pi(C)=const.$ and $\Pi(I)=const.$ 
This means that the posterior probability is simply equivalent to the likelihood function. 
The posterior probability is normalized so that the total is unity. 
The $1\sigma $, $2\sigma $, and $3\sigma $ contours of the posterior probability are close to those of $\Delta \chi^2$ measure \citep[e.g.,][]{Snellen+2017}, with a slight difference due to the non-linear $I$-dependence of the model.

\subsubsection{Analysis (B): with unknown stellar spectrum}
\label{sss:analysis_pessimistic}

In the second analysis, we do not assume that the stellar spectrum is known a priori. 
Instead, we consider the time average of the combined spectra of the star and the planet and assume that the stellar spectrum is fully included in this averaged spectrum, which is valid if the stellar spectrum is stable in time. 
On the other hand, the spectral features of the planet should remain in the average-subtracted residual spectrum, due to the Doppler shift caused by its orbital motion. 

The analysis procedure is as follows:
\begin{enumerate}
    \item The average of the spectra at different planetary orbital phases is obtained. 
    \item The average spectrum is subtracted from the spectrum at different orbital phases.  
    \item The model planet spectrum, which is also Doppler-shifted and average-subtracted (the bottom panel of Figure \ref{fig:schematics2}), is fitted to the set of residual spectra. 
\end{enumerate}
The parameter estimate in Step 3 is performed in the same way as Analysis (A) (equations (\ref{eq:postprob})-(\ref{eq:chi2})).

Figure \ref{fig:schematics2} illustrates the signal that can be used in this analysis. 
The average of the spectrum at $\phi = 90^{\circ }$ and that at $\phi = 270^{\circ }$ is shown in the black line in the upper panel, and the average-subtracted spectra are shown in the lower panel. 
During this subtraction process, 
some portions of the planetary spectral features are cancelled out, reducing the signal level. 
This effect is substantial in the case of MIR observations of potentially habitable planets whose strongest lines are broader than the Doppler shift (Figure \ref{fig:HRspectra}). 
The situation is different from the typical high-resolution spectroscopy of hot Jupiters at the shorter wavelengths, where the lines are narrower and the Doppler-shift is larger. 
In such a case, the Doppler shift moves the line beyond their intrinsic width and the average-subtracted spectra are more similar to the original spectrum.

\subsubsection{Notes on cross-correlation analysis}

High-resolution spectroscopy of hot-Jupiter systems have been routinely analyzed through cross-correlation function \citep{Snellen+2010,Brogi+2013}. 
While this technique is useful for detecting high-frequency features, its statistical treatment and the uncertainties in the estimated model parameters are not straightforward \citep[e.g.,][]{Brogi+2017}. 
On the other hand, fitting observed data with theoretical models by minimizing the sum of squared residuals offers a more plain interpretation \citep{Snellen+2017}. 
Thus, in this paper, we employ the latter method.

\subsection{Results}
\label{ss:results}

\begin{figure*}[tb!]
\includegraphics[width=\hsize]{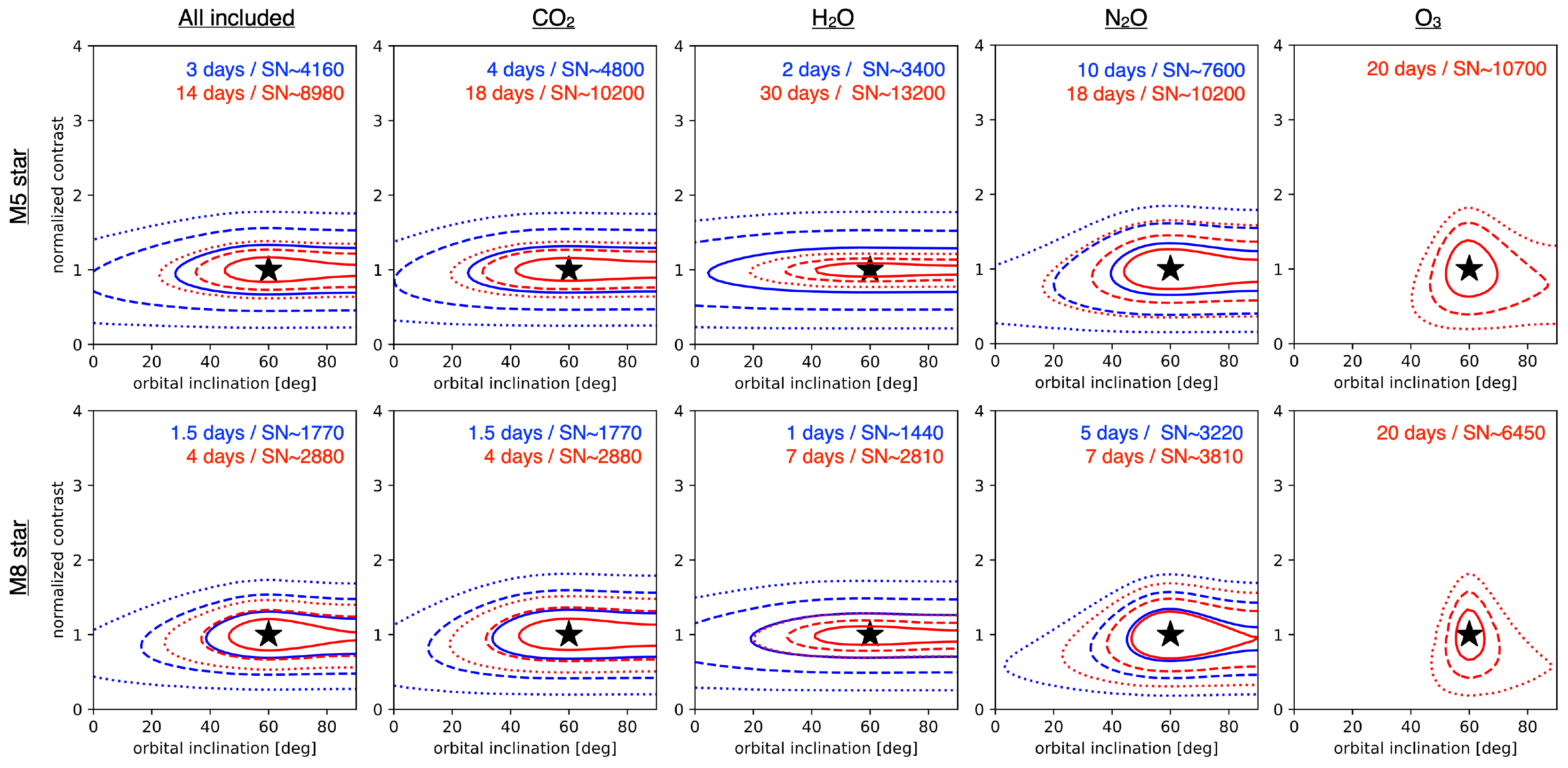}
\caption{The 1$\sigma $ (solid lines), 2$\sigma $ (dashed lines), and 3$\sigma $ (dotted lines) confidence intervals for planets around a mid-M star (upper panels) and a late-M star (lower panels), based on analysis (A), i.e., when the stellar spectrum is precisely determined. Both the mock data and the fitting model are based on the model thermal emission spectra presented in Figure \ref{fig:HRspectra}. The constraints from two integration times are presented for some cases, in order to show how the constraints are developed as the integration time becomes longer. The blue colors imply that the presence of the molecule is detected (i.e., the $C=0$ is rejected by 3$\sigma $) and the inclination is not well constrained, while the red colors imply that both the contrast and inclination are constrained. The corresponding signal-to-noise ratios of the host star spectrum per wavelength resolution element (SN) are also reported. }
\label{fig:postprob_A}
\end{figure*}

\begin{figure*}[tb!]
\includegraphics[width=\hsize]{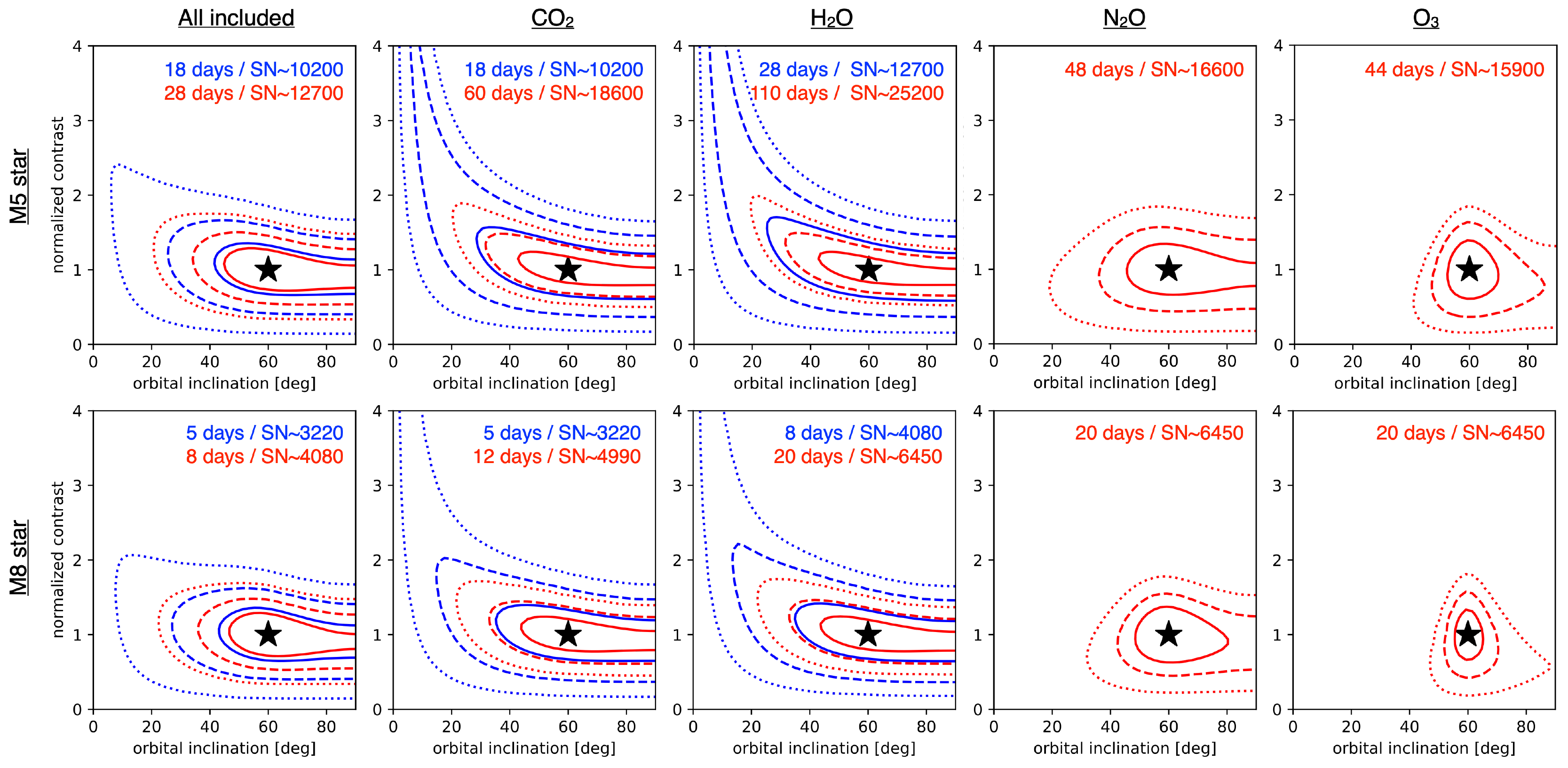}
\caption{Same as Figure \ref{fig:postprob_A}, but the mock data are now analyzed by analysis (B).}
\label{fig:postprob_B}
\end{figure*}

Figures \ref{fig:postprob_A} and \ref{fig:postprob_B} summarize the results of Analysis (A) and Analysis (B), respectively, performed on the thermal emission spectra with all molecules included and with individual molecules in isolation (Figure \ref{fig:HRspectra}). 
The solid, dashed, and dotted lines present the 1$\sigma$, 2$\sigma$, and 3$\sigma $ constraints in the contrast-inclination plane (i.e., 68.27\%, 95.45\%, and 99.73\% probability), after the indicated integration time. 
The corresponding average signal-to-noise ratio (SN) of the host star spectra per wavelength resolution element is also presented.

As expected, detecting any molecule is easier for planets around late-M stars than around mid-M stars due to the larger planet-to-star flux ratio and, for analysis (B), to the larger radial velocity amplitude. 
In the model considered here, the total observation time required for a M5-star system is by a few times larger than a M8-star system, or 3 times smaller signal-to-noise ratio of the host star. 

When the host star spectrum is accurately known and corrected (Analysis (A); Figure \ref{fig:postprob_A}), 
the spectral features of an Earth-like atmosphere at 5 prsecs away can be detected within a few days of total observation time, even around mid-M systems (the leftmost column). 
Compared to this optimistic scenario, Analysis (B) requires approximately 4 times longer observations to detect the spectral features of the same target, or doubled signal-to-noise ratio of the host star flux. 
This is because of the self-subtraction discussed in Section \ref{sss:analysis_pessimistic}.

For comparison, we also perform the same analysis assuming only one of the molecules as opacity source (the lower five panels of Figure \ref{fig:HRspectra}); the results of which are shown in the right four columns of Figures \ref{fig:postprob_A} and \ref{fig:postprob_B}. 
Comparison of the two figures clearly shows that the effect of the self-subtraction is substantial except for O$_3$. 
In addition, for Analysis (B), the contrast and the inclination angle tend to degenerate for pure CO$_2$ or pure H$_2$O atmospheres. 
This is because when a small orbital inclination (i.e., close to face-on) is assumed, the Doppler shift of the planetary spectrum is small and the amplitude of the differential spectrum (the lower panel of Figure \ref{fig:schematics2}) is also small, which is then compensated by a large contrast. 
Note that the orbital inclination is poorly constrained even with Analysis (A) for pure CO$_2$ or pure H$_2$O atmospheres, due to the broad nature of the spectral lines.

Compared to these two molecules, the self-subtraction is not substantial for O$_3$ because the O$_3$ bands are densely populated by narrow lines, and the detectability through Analysis (B) is similar to that of Analysis (A). 
Furthermore, the orbital inclination is well constrained as soon as the contrast is constrained to non-zero. 
N$_2$O features are also relatively good at constraining the inclination, despite its relatively weak features. 

In the spectrum with all the molecules included, the features of each molecule tend to be more muted than the features of individual molecules in isolation, due to the line overlaps. 
However, the spectrum become more rich in features, and the constraints on the contrast is as good as for the pure-CO$_2$ or pure-H$_2$O cases.
Furthermore, the overlaps of the spectral features of individual molecules break the degeneracy between the contrast and inclination for Analysis (B), significantly improving the constraints on the orbital inclination compared to the pure-CO$_2$ or pure-H$_2$O cases. 
As a result, both the contrast and the orbital inclination can be reasonably constrained within $\sim 1$ week of integration for M8 star systems within 5~parsecs even with Analysis (B). 
For earlier type systems, constraining the orbital inclination would become more challenging.

\subsubsection{Scaling and the application to the known targets}
\label{sss:scaling}

While the estimated integration time would be too long to be practical except for the case of late-M stars, 
it is possible to approximately scale these numbers for varying observational configurations. 
When the stellar light is the dominant noise source, the integration time that achieves certain noise level, $\tau_{\rm esp,\,0} $, is proportional to:
\begin{equation}
    \tau_{\rm esp,\,0}  \propto \left( \frac{R_p}{R_{\oplus }} \right)^{-2} \left( \frac{d}{\mbox{5 pc}} \right)^2 \left( \frac{D}{\mbox{6.5 m}} \right)^{-2} \left( \frac{\xi}{0.2} \right)^{-1} \label{eq:scaling}
\end{equation}
In particular, the observations of the nearest possible target, Proxima Centauri b, at the distance of 1.3 pc and with the estimated radius of $\sim 1.1 R_{\oplus }$, would only require $\sim $1 days of observation even without an assumption on the stellar spectrum. 

A more SPICA-like specification with $D=2.5$~m and a lowered total throughput ($\xi=0.1$) yield $\sim $14 days of total integration time.

\section{Dependence of Detectability on Spectral Resolution, Molecular Abundance, and Bandpass}
\label{s:Detecability_Resolution_Abundance}

In the assessment in Section \ref{s:detectability}, we have made several assumptions for the specifications of the spectrographs and the atmospheric profile. 
In this section, we discuss the effects of varying these assumptions on the detectability of the molecules. 

\subsection{Effects of Spectral Resolution}
\label{ss:optimal_resolution}

First, we discuss the dependence on the spectral resolution of the spectrograph. 
Let us represent the signal by the total area of the shaded region shown in Figure \ref{fig:schematics2}, namely the integration of the absolute value of the differential spectrum both in Analysis (A) and (B); For Analysis (A), the differential spectrum corresponds to the difference between the original spectrum and the moving average, while for Analysis (B) it is the difference between the spectrum at a certain orbital phase and the time-averaged spectrum. 

\begin{figure}[tb!]
\includegraphics[width=1.0\hsize]{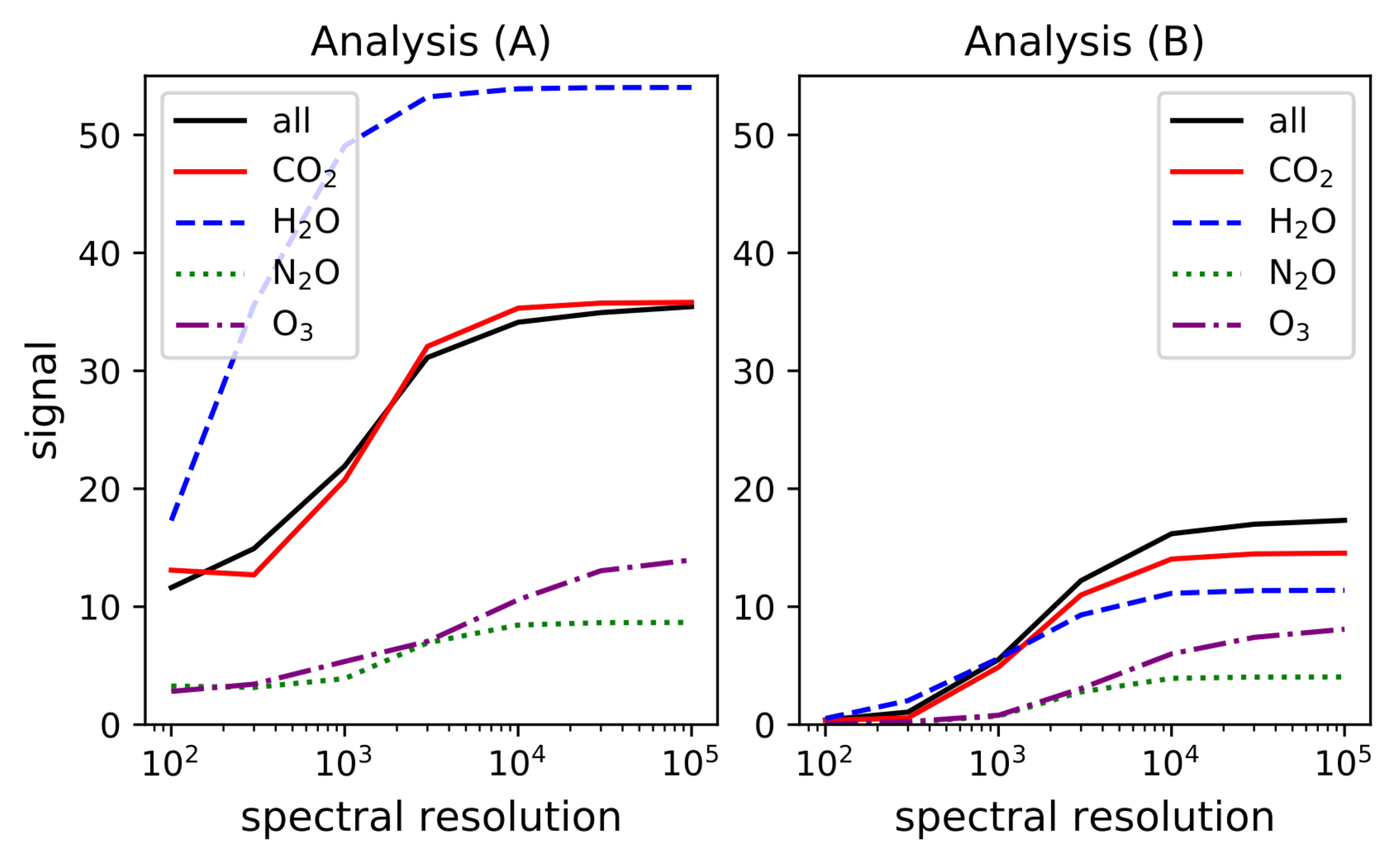}
\caption{The total areas of the filled region in Figure \ref{fig:schematics2}, one of the diagnostics of the signal, as a function of the spectral resolution, using Analysis (A) (left) or Analysis (B) (right). The fiducial atmospheric profile is assumed. }
\label{fig:S_vs_R_fiducial}
\end{figure}

\begin{figure}[tb!]
\includegraphics[width=1.0\hsize]{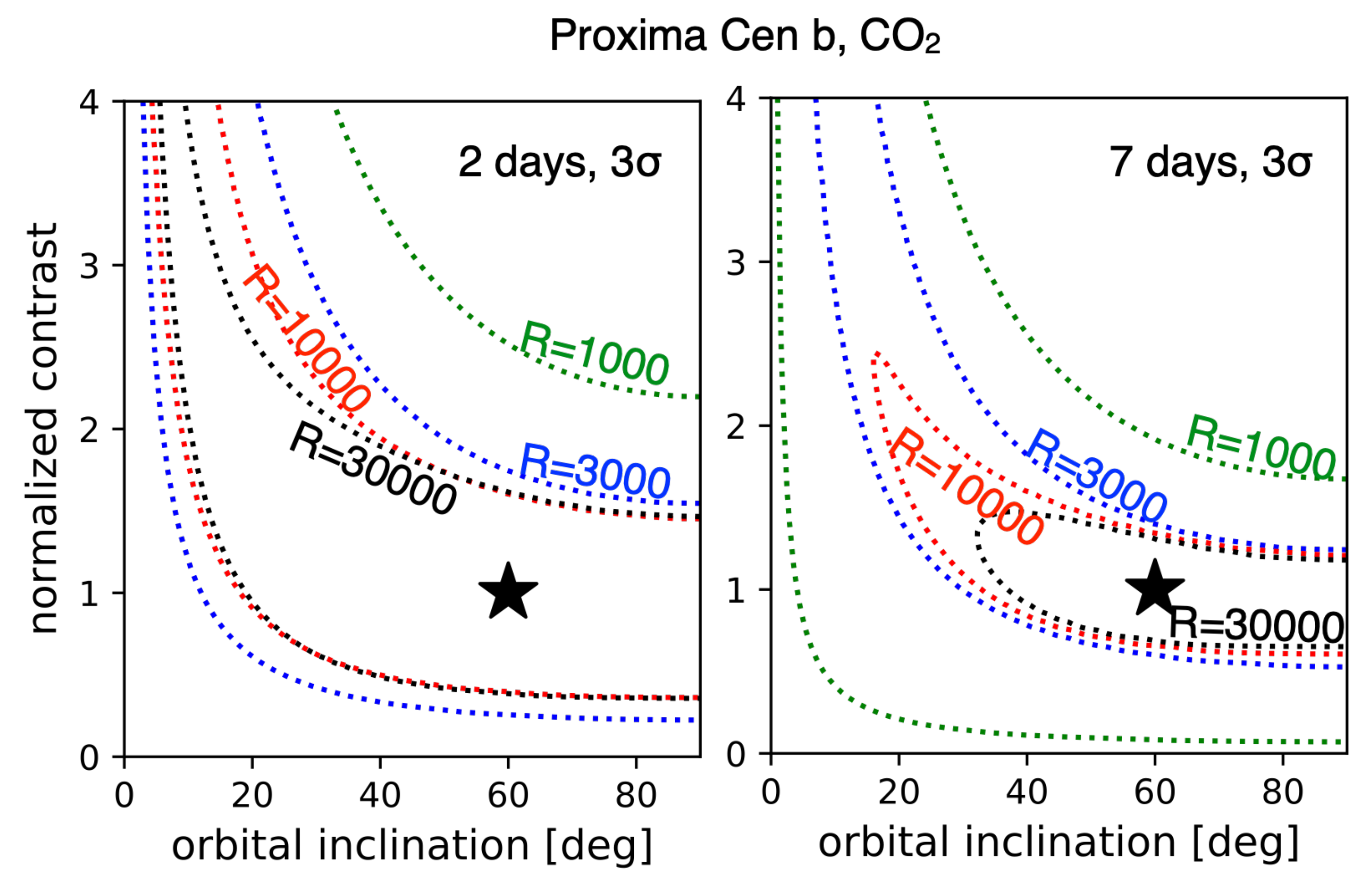}
\caption{3$\sigma $ constraints in the contrast-inclination plane based on the observations of Proxima Centauri b with CO$_2$ features using varying spectral resolution, after 2 days (left) or 7 days (right) of integration. }
\label{fig:ProxCenb_CO2_varyR}
\end{figure}

Figure \ref{fig:S_vs_R_fiducial} shows the signals defined above for Analysis (A) and (B) as a function of the spectral resolution. 
Black line shows the case of the spectrum with all four molecules included (the bottom panel of Figure \ref{fig:HRspectra}), while other lines show the cases for the spectra of individual molecules. 
As expected, the signal level monotonically increases as the spectral resolution increases. 
However, it is mostly saturated above $\mathcal{R} \gtrsim 10,000$ except O$_3$, in both analysis (A) and (B). 
This is partly because the interval of the strongest lines are $\sim 0.1-0.2\,\mu $m and these lines start to be resolved with $\mathcal{R} \gtrsim 2000$. 
In addition, the intrinsic line broadening starts to dominate over the broadening determined by the spectral resolution for $\mathcal{R} \gtrsim 10,000$.
The required resolution, $\mathcal{R} \gtrsim 10,000$, is about one order of magnitude smaller than that in the visible or near-infrared domain ($\lambda \sim 1\,\mu$m), $\mathcal{R} \gtrsim 100,000$, because the photon energy is about one order of magnitude lower while the energy corresponding to the line broadening and the interval of lines do not depend on the wavelengths (See the discussion in Section \ref{ss:result_spectra}). 

The saturation is less prominent for O$_3$, where the narrow lines are densely populated. 
Thus, detection O$_3$ benefits from the increased spectral resolution beyond $\mathcal{R}\sim 10,000$. 

Figure \ref{fig:ProxCenb_CO2_varyR} shows the constraints for the contrast and the orbital inclination obtained through Analysis (B) of mock observations of Proxima Centauri b for 2 days (left) and 7 days (right) with varying resolution. 
As expected from the above assessment, the detectability, namely the lower constraints on the contrast, improves as the spectral resolution increases up to $\mathcal{R}\sim 10,000$, above which no significant improvement is seen. 
However, the constraints on the orbital inclination is improved beyond $\mathcal{R}\sim 10,000$.

\subsection{Choice of the bandapss}
\label{ss:choise_of_bandpass}

Our fiducial bandpass is set to 12-18 $\mu $m, based on the planned specification of SPICA/SMI and OST/MISC. 
In this range of wavelengths, several other molecules have prominent absorption bands, in addition to the molecules studied in this manuscript. 
They includes NH$_3$, SO$_2$, and NO$_2$ and they may also be useful for characterizing the surface environment of potentially habitable planets. 

Although there is no planed high-resolution spectrograph that goes beyond this bandpass, it would be worth considering the impact of the alternative choice of bandpass for future instrumental designs. 
In the following, we briefly discuss how the extension of the bandpass would affect the prospects. 

If the bandpass is extended toward shorter wavelengths, a feature of particular interest would be O$_3$ 9.7~$\mu $m band (the bottom panel of Figure \ref{fig:HRspectra}). 
The peak line strength of this band is approximately one order of magnitude larger than the 14.5 $\mu $m band and, combined with the fact that the shorter wavelength is in the Wien's regime of the Planck function, the feature relative to the continuum is deeper. 
However, the reduced planet-to-star contrast (Figure \ref{fig:contrast-photoncount}) does not make this band easy to detect. 
We find that detection of O$_3$ 9.7 $\mu $m O$_3$ band would require approximately 1.7 times longer integration time (or 75\% photon noise) than the detection of 14 $\mu $m O$_3$ band, in the absence of CO$_2$ (not shown). 
In the presence of CO$_2$ larger than 1 ppm, 9.7 $\mu $m band would be the only detectable band. 
The detectability of O$_3$ 9.7 $\mu $m would be improved for planets with higher surface temperature. 
The features at even shorter wavelengths is even less likely to be detectable with this method due to the small planet-to-star contrast. 

At the wavelengths longer than 18 $\mu $m, there are few vibrational modes of simple molecules. 
However, the rotational lines of H$_2$O extend beyond 18 $\mu $m and adding longer wavelengths improves the detectability of H$_2$O. 
Using 12-24 $\mu $m bandpass, the integration time required for the detection of H$_2$O from the H$_2$O-only spectrum (the second panel of Figure \ref{fig:HRspectra}) would be reduced to approximately 60\%.

\subsection{Molecular abundances that can be most sensitively probed}
\label{ss:optimal_abundance}

The detectability also depends on the actual abundance of the  molecules. 
While in transmission spectroscopy the detectability of certain molecules generally increases as the abundance increases (ignoring the effects of clouds), this is not the case for the high-resolution method considered in this paper. 
It is true that the signal would not be detected if the abundance is too small. 
However, larger abundances does not always lead to the improved detectability, because some of the features start to saturate. 
For example, with the assumed Earth-like abundance of CO$_2$ ($\sim $ 330~ppm) the strongest lines of CO$_2$ around 15~$\mu $m are saturated (i.e., the spectrum is flattened) and do not significantly contribute to the signal. 
With larger abundance, the flattened region only increases, although the features associated with smaller opacities kick in. 
Thus, there is the optimal range of molecular abundance.

\begin{figure}[tb!]
\includegraphics[width=1.0\hsize]{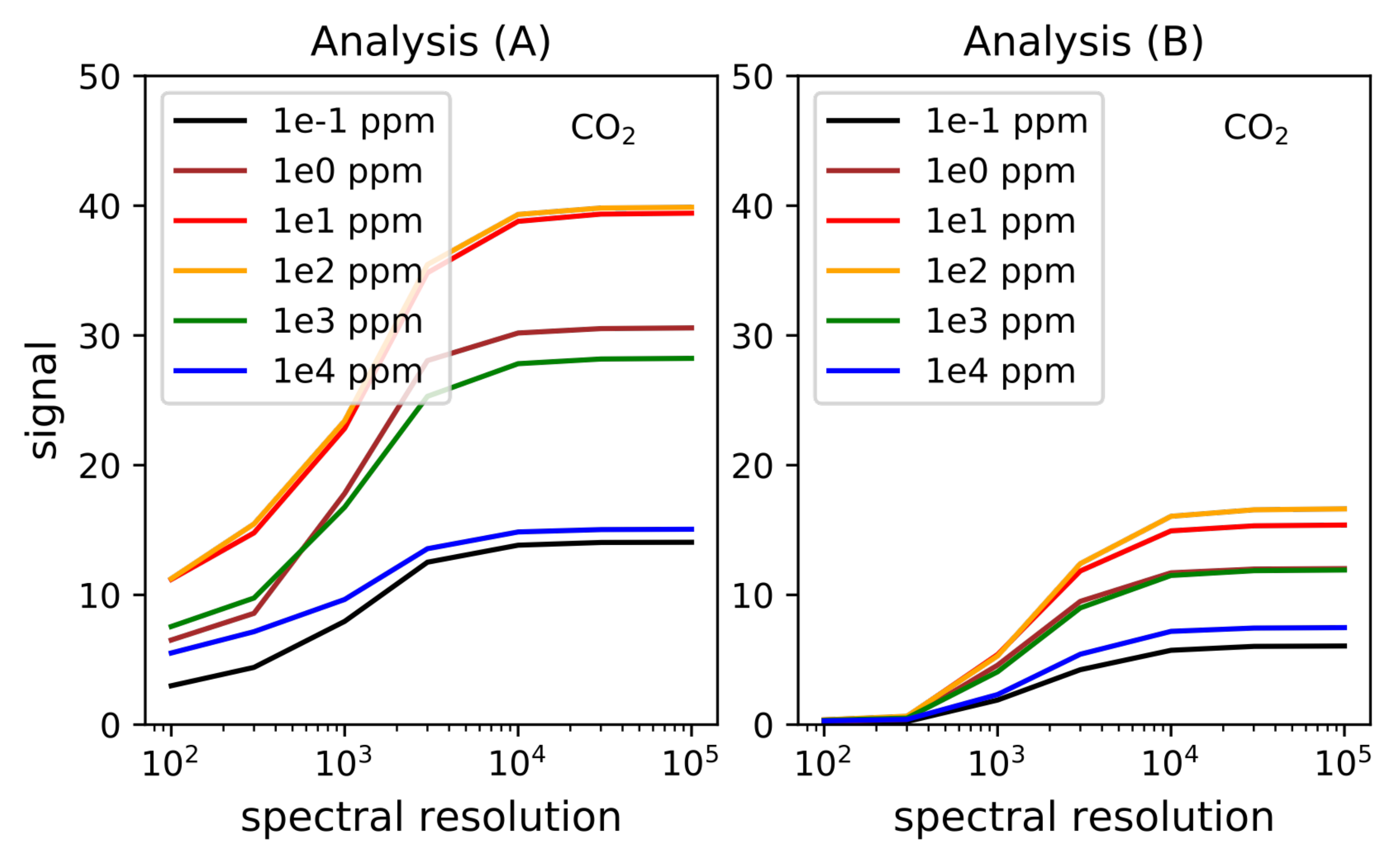}
\caption{Same as Figure \ref{fig:S_vs_R_fiducial} but with varying mixing ratio of CO$_2$.}
\label{fig:S_vs_R_CO2}
\end{figure}

To see such a dependence of detectability of CO$_2$ and N$_2$O, we created model spectrum with varying mixing ratio assuming a 1-bar atmosphere with vertically uniform mixing ratio of each molecule. 
For each model spectrum, we repeated the analysis of Section \ref{ss:optimal_resolution} to see the dependence of the ``signal'' as a function of the spectral resolution. 
The result is shown in Figure \ref{fig:S_vs_R_CO2}. 
The signal level is maximized when the mixing ratio is 1-10$^3$~ppm. 
A similar trend was found for N$_2$O, with the signal level being maximized for 1-10$^2$~ppm.  
Therefore, this technique is sensitive to relatively small amount of molecules. 
This would be rather complementary to scattered-light observations where larger abundance of these molecules would be probed; see Section \ref{ss:targets}.

\section{Discussion}
\label{s:discussion}

\subsection{Potential Targets and Synergy with other techniques}
\label{ss:targets}

The prime targets of the method studied in this manuscript would be potentially habitable planets around mid- to late-M stars within 5~parsecs. 
The fact that this method does not require planetary transits is a significant advantage over transmission or eclipse spectroscopy, given the low transit probability ($\sim $5\%). 
Within approximately 5~parsecs, 6 mid- to late-M stars have already been found to harbor Earth-sized planets around habitable zones as of the submission of this manuscript (Table \ref{tbl:targets}), none of which are found to transit in front of the host star. 
The high-resolution technique would provide valuable opportunities to investigate the atmospheric compositions of these known planets in thermal emission.
The number of the targets may increase thanks to the current vigorous radial velocity measurements in near-infrared all over the world (There are 7 late-M stars (M6 or later) and 10 mid-M stars (M3-M5) within 5~parsecs, according to \citet{Gaidos+2014}). 
Note that we do not necessarily limit our scope to 5 parsecs. Some of the features that are easier to detect may extend its scope beyond 5 parsecs. 

These nearby M-star systems are also considered to be the prime targets for the scattered-light observations with the next-generation extremely-large ground-based telescopes \citep{Fujii+2018}. 
The combined efforts would allow for the comprehensive characterization of the target systems, as the information obtained from scattered-light spectra (e.g., surface liquid water, clouds, certain molecules) and that obtained from thermal emission spectra (temperature, radius, certain molecules) are complementary. 
For example, thermal emission spectra could in principle constrain the planetary radius, which critically helps the interpretation of the scattered-light spectrum. 
Furthermore, for a given molecule, the abundance detectable in the scattered light at $\sim $1~$\mu $m would be larger than the abundance probed by the high-resolution technique (Section \ref{ss:optimal_abundance}), thus the combination would constrain a wide range of atmospheric properties.  

The targets of MIR high-resolution technique are also complementary to the prime targets for transmission and eclipse spectroscopy. 
The main scope of OST is currently to characterize potentially habitable planets through transit spectroscopy by achieving ultrahigh stability (see the technical report of OST). 
There are pros and cons for transit spectroscopy and for the technique studied in this paper. 
While the former may require less total observation time for certain biosignature molecules (provided that the noise floor could be down to a few ppm) than the latter, it has to visit the target 50 times for TRAPPIST-1 planets \citep{Tremblay+2020} and more for less favorable targets, likely spanning ~1 year or longer. 
The number of visits of the high-resolution method studied in this manuscript can be much smaller because of the longer observation time per visit.  
In addition, the transmission spectroscopy could be hampered  by high-altitude thin clouds, which does not significantly affect the high-resolution technique. 
Thus, combining both techniques would expand our capability of investigating potentially habitable worlds.

\begin{table}[tb!]
\centering
\caption{Nearby systems that are found to harbor Earth-sized planets around habitable zones. Stellar parameters are based on \citet{Gaidos+2014}. }
\begin{tabular}{lccc}\hline \hline
Star name & $d$ & Spectral type & $T_{\star}$  \\ \hline
Proxima Cen & 1.3~pc & M7 & 2883~K \\
Ross 128 & 3.4~pc  & M5 & 3145~K \\
Gliese 1061 & 3.7~pc & M6? & 3000~K? \\
Luyten's & 3.8~pc & M4 & 3317~K \\ 
Teegarden's & 3.9~pc & M7 & 2700~K \\
GJ 682 & 5.1~pc & M5 & 3190~K \\ \hline
\end{tabular}
\label{tbl:targets}
\end{table}

\subsection{Effects of other noise}
\label{ss:other_noise}

\subsubsection{Time variability of stellar spectra}
\label{sss:stellar_activity}

The major concern of this method is the variability of high-resolution stellar spectra. 
Aside from sporadic flares, stars generally exhibit photometric and spectroscopic variabilities mainly due to starspots.
Given that the signal from the planet is a tiny fraction of the stellar spectrum, such variability potentially cases serious problem in the analysis. 

There is a growing amount of modeling efforts for the temperatures and covering fractions of star spots/faculae of mid- to late-M stars based on observations. 
One piece of evidence comes from Doppler imaging. 
Doppler imaging of mid- to late-M stars  \citep{Morin+2008,Barnes+2015,Barnes+2017} estimates that around a few \% of the surface is covered with spots that are assumed to be cooler than other area by 200-400~K. 
Recently, characterization of TRAPPIST-1, an M8 star harboring transiting temperate Earth-sized planets, has attracted attention.
The light curve of TRAPPIST-1 observed by Kepler/K2 (0.43-0.89 $\mu $m) shows $\sim 1$\% peak-to-peak variation amplitude, which may be explained by rotating dark spots or faculae. In contrast, the variability of {\it Spitzer} at 4.5 $\mu $m do not show a clear variability and is at least smaller than the photometric precision, which is limited to the order of 0.1\% owing to the shot noise and instrumental systematic noise. 
\citet{Morris+2018} find that the combination of these light curves at different wavelengths can be explained by the existence of several very small faculae with $> 5300$~K rather than dark spots, and show that the theoretical photometric variability at 4.5~$\mu $m with this model seems to be random with an amplitude of about 100~ppm. 
This prediction is encouraging because this is compatible or smaller to the contrast of the planet to stellar signals. 
The impact of variability on this method will be further mitigated by averaging the data over various stellar rotational phases.

The covering fraction of the spots and faculae are also estimated based on the multi-component fit to the stellar spectrum using the theoretical stellar spectra with various temperatures  \citep[e.g.,][]{Zhang+2018,Wakeford+2019}. 
They typically result in relatively large covering fractions for darker spots and faculae (about 30-50\%). 
However, this method is sensitive to the small-scale features that spread across the photosphere as well as the large-scale asymmetric patterns, in contrast to the Doppler Imaging and reconstruction from photometric variabilities that are only sensitive to the latter. 
The small-scale features that distribute rather uniformly do not affect the observations studied in this manuscript.

We note that the recent observations of Proxima Centauri by \citet{SuarezMascareno+2020} show V-band photometric variability of $>$10\% (peak-to-peak), larger than the {\it Kepler/K2} variability of TRAPPIST-1. 
In reality, stars exhibit the diversity in their surface magnetic activities and the resultant spots/faculae, and therefore it is critical to characterize individual stars before characterizing the planets. 
With the launch of JWST that achieves higher precision than {\it Spitzer} thanks to its larger aperture and better attitude control, our understanding of variabilities of the stellar MIR spectra will be developed and the techniques to cope with these variability will be advanced. 
For example, simultaneous light-curve observations at different wavelengths will allow us to better model the spots, as demonstrated by \citet{Morris+2018}. 
Such practice will further enhance the feasibility of the method presented in this manuscript.

\subsubsection{Systematic Noise}
\label{sss:systematic_noise}

We did not explicitly deal with the impact of the unknown instrumental systematic errors. 
Because this technique utilizes high-frequency features of the spectrum, the uncertainties in the absolute flux or the low-frequency modulation of the continuum, even if it is substantial, will not affect our results after applying some corrections \citep[e.g., high-pass filter;][]{Snellen+2017}.

In contrast, residual flat-fielding, spectral fringe patterns, and telescope pointing jitters form artificial absorption and emission lines (i.e., high-frequency systematic noise) in the observed spectrum. Such artificial high-frequency patterns will be problematic for both the analyses (A) and (B). 
The high-frequency noise will contribute to the total noise floor because of its random distribution along the wavelength.  Since the high-frequency systematic noise is independent of the shot noise, the total noise may be expressed by $\sqrt{ \sigma^2_{{\rm photon},j} + \sigma^2_{{\rm sys},j} }$, where $\sigma_{{\rm photon},j} $ is defined by equation (\ref{eq:sigma}) and $\sigma_{{\rm sys},j}$ denotes the high-frequency systematic noise at the $j$-th spectral element. 
Given that the noise required for detection is order of 100~ppm, it is necessary that the instrument systematic noise is less than this level. If the systematic noise is $\sim 30$~ppm or less, the systematic noise will not affect our estimate much. 
Conversely, if it is comparable to the photon noise, a substantially larger integration time will be required in order to reduce the photon noise further. We note that the high-resolution spectroscopy observes the shift of the planetary Doppler signal on the detector, which could mitigate an impact of the patterns fixed to the detector (owing to relative sensitivity variations between adjacent spectral channels).

\subsection{Variety of Planet Thermal Emission Spectrum}
\label{ss:dependence_on_atmosphere}

\subsubsection{Stratospheric Thermal Inversion}
\label{sss:thermal_inversion}

\begin{figure*}[tb!]
\includegraphics[width=1.0\hsize]{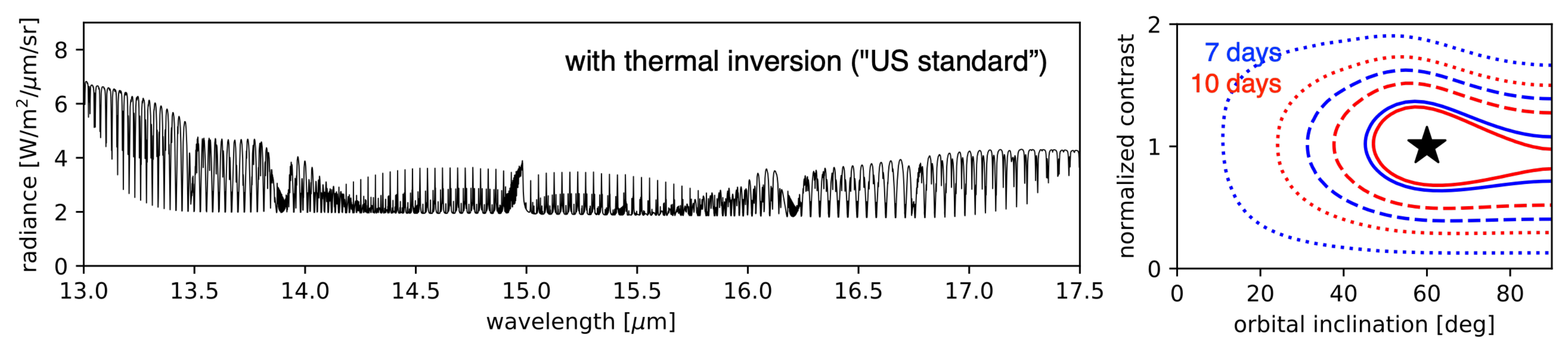}
\caption{Left: high-resolution ($\mathcal{R}=30,000$) thermal emission spectra assuming the vertical profiles of temperature and CO$_2$ that are same as Earth's ``US standard'' profile. Right: Parameter constraints based on the left panel, assuming a late-M host star at 5 parsec. }
\label{fig:sp_postprob_USS}
\end{figure*}

This section discusses the variety of high-resolution thermal emission spectra, and how they affect the parameter estimate. 

Our fiducial model atmosphere has a relatively cool stratosphere, approximately 150~K. 
A higher stratospheric temperature results in shallower spectral features, increasing the required integration time for detection. 

If the planet has thermal inversion in the upper atmosphere (like the ozone layer of Earth), thermal spectra may exhibit emission lines. 
These lines can be sharp, because the pressure in the upper atmosphere (where the emission line are formed) is low and the line width is relatively narrow. 
This can relax the degeneracy between the contrast and the orbital inclination which would otherwise be present, like our fiducial cases for CO$_2$ and H$_2$O. 

As a demonstration, Figure \ref{fig:sp_postprob_USS} is the modeled thermal emission spectrum that assumes an Earth-like temperature profile with a stratospheric thermal inversion as well as an Earth-like CO$_2$ vertical profile. 
Note the narrow emission lines within 14-16~$\mu $m. 
The analysis (B) performed on this spectrum yields the constraints shown in the right panel of Figure \ref{fig:sp_postprob_USS}, assuming a late-M host star. 
While the total integration time required for detection is longer than our fiducial case due to the reduced line depths (corresponding to the warmer upper atmosphere), the degeneracy between the contrast and the inclination is less severe. 

\subsubsection{Horizontal temperature gradient}
\label{sss:full_exploration}

We did not include the horizontal temperature gradient of the planet. 
In reality, it can be substantial near the surface, depending on the heat re-distribution efficiency of the atmosphere and/or ocean. 
The horizontal temperature gradient would result in the time variation of the disk-averaged thermal emission as the planet rotates \citep[e.g.,][]{Knutson+2007}. 
While this itself can be used to characterize planet atmosphere, such variations may complicate the analysis of high-resolution spectroscopy. 
Fortunately, the effect of the horizontal temperature gradient will be mitigated if the data at the same phase angle are stacked and analyzed.  
Our analysis assumes observations near 90 degrees and 270 degrees (i.e., the same phase angle) so the impact will be limited unless there is an extreme asymmetry between the eastern and western hemispheres.

\subsubsection{Full Exploration of Parameter Space}
\label{sss:full_exploration}

Although in this paper we attempted to constrain the contrast and inclination by fitting the mock data by specific model spectra, in reality, the spectral shape of the planet thermal emission also has to be constrained. 
In other words, the distribution of the molecules as well as the temperature structure should be fitting parameters as well, and the dimension of the parameter space is large. 
It is also possible that the constraints on certain parameters can be affected by the prior probability when exploiting the Bayesian framework. 
Full explorations of the parameter space and development of sophisticated retrieval techniques will need further study.  
Also, our analysis implicitly assumes that the parameter space explored includes the ground truth. If this is not the case, the obtained constrains should be interpreted with a special care, which is to be examined in future studies.

\subsection{Significance of mid-infrared high-resolution spectroscopy for characterization of potentially habitable planets}
\label{ss:comp_MRS}

Because our analysis (A) follows the procedure of \citet{Snellen+2017}, we compare our results with their estimate. 
Their modeled CO$_2$ features in the medium-resolution spectra is detected (by nearly 4 $\sigma $) after the photon noise becomes $\sim $50~ppm. 
Using the same bandpass (13.2-15.8 $\mu $m) and a similar temperature profile (left panel of Figure \ref{fig:sp_postprob_USS}), our analysis (A) of high-resolution spectra was able to detect the CO$_2$ features when the photon noise is $\sim $100~ppm. 
This is likely a consequence of about two times sharper features at high-resolution spectra, as well as the minor features that are smoothed in the medium resolution spectra. 

Note that it is not trivial to compare our estimate for the integration time to the integration time estimated in \citet{Snellen+2017} for JWST/MIRI. 
This is because the noise of JWST/MIRI, especially beyond 12~$\mu $m, is not dominated by the photon noise of the star. 
Other noise, including the thermal background of the telescope, zodiacal light (with the larger aperture size), and/or the readout noise (due to the short exposure time assumed) suppresses S/N per exposure by 2-3 times compared to the idealized case. 
This leads to the substantial difference in the integration time required to achieve a certain precision. 
Thus, it is critical to suppress the background noise in order to be able to detect atmospheric molecules of nearby mid-M and late-M stars in a reasonable amount of time.

\section{Summary}
\label{s:summary}

In this paper, we study the use of an MIR high-resolution spectrograph mounted on a cryogenic telescope for characterizing non-transiting temperate terrestrial planets orbiting M-type stars.
We modeled high-resolution thermal emission spectra of an Earth-like atmosphere with CO$_2$, H$_2$O, N$_2$O, and O$_3$ assuming a simplified temperature profile composed of an dry adiabatic troposphere and an isothermal stratosphere.  
We show that the MIR spectral features of Earth-like atmospheres can be broader than the width of the Doppler shift, depending on the atmospheric structure. 

In order to reasonably identify the tiny planetary features in the combined spectra of the star and the planet, it is critical to subtract the stellar contribution precisely, as low-mass stars are particularly rich in spectral features in the MIR. 
For non-transiting planets, the stellar spectrum cannot be determined observationally as we always observe the star and planet together. 
Considering this possible difficulty, we proposed to observe the target at around $\phi = 90^{\circ}$ and at $\phi = 270^{\circ}$ and that the differential spectra are fitted by the model. 
This process can reduce the signal because some fraction of broad lines are self-subtracted. 
This effect is substantial except for O$_3$, increasing the total integration required for detection by several times.

Nevertheless, the spectral features ($\mathcal{R}=30,000$) of an Earth-like planet between 12-18 $\mu $m would allow us to constrain the contrast and the orbital inclination within $\sim $1 days of total integration time, for an M8-star system within 5~parsecs. 
For earlier-type stars, constraining the orbital inclination may be challenging. 
Scaling these estimates through to Proxima Centauri b, an Earth-like abundance of CO$_2$ would be inferred in $\sim $1 day of integration. 
In addition to Proxima Centauri b, several potentially habitable Earth-sized planets have been already discovered within 5 parsecs, and they are the prime targets for this technique. 

We find that the constraints on the contrast improves with higher spectral resolution, but $\mathcal{R} \gtrsim 10,000$ do not result in significant improvement except for O$_3$ whose absorption bands are densely populated with narrow lines. 
However, the constraints on the inclination would benefit from the higher spectral resolution. 
We also find that this method is most sensitive to relatively small amount of CO$_2$ and N$_2$O (1-10$^{3}$~ppm for a 1-bar Earth-like atmosphere), where the higher abundance does not lead to better detectability.

In this study, we did not include the stellar variability and systematic noise in our simulation. 
Because the photon noise required for the detection of these features is 100~ppm or larger, 
the stellar variability and systematic noise must be suppressed to these levels in order to detect the planetary features. 
The expected stellar variability in the mid-infrared wavelength range based on the previous photometric observations of TRAPPIST-1 with {\it Spitzer} is comparable or less than 100 ppm, which will not affect this method. Suppression of the noise sources other than the shot noise from the stellar flux (e.g., thermal background) is critical in reducing the required integration time and making the observations realistic. 

\acknowledgments
We thank Teruyuki Hirano for helpful discussions on high-resolution spectroscopy and Klaus Pontoppidan for the information about the instrumental noise of JWST. 
YF is supported by Grand-in-Aid from MEXT of Japan, No.~18K13601. 
TM is supported by Grand-in-Aid from MEXT of Japan, No.~19H00700.

\bibliography{ref}

\end{document}